\documentclass[prd,aps,preprint,tightenlines,superscriptaddress,nofootinbib,showpacs,showkeys]{
 revtex4}
\usepackage{amsfonts}
\usepackage{array}
\usepackage{epsfig}
\usepackage{rotating}
\usepackage{multirow}

\usepackage{amsmath}    
\usepackage{verbatim}   
\usepackage{color}      
\usepackage{colordvi}

\usepackage{subfigure}  
\usepackage{hyperref}   

\usepackage{pstricks,rotating, graphicx}

\graphicspath{{../pics/}}
\DeclareGraphicsExtensions{.eps,.ps}

\usepackage{pslatex}
\usepackage{txfonts}
\usepackage[latin1]{inputenc}
\usepackage[T1]{fontenc}

\newcommand{\GeV}{\ensuremath{\,\mathrm{GeV}}}

\newcommand{\msbar}{$\overline{\mathrm{MS}}\, $}

\begin{document}

\begin{titlepage}
\thispagestyle{empty}
\noindent
DESY 20--061
\hfill
\\
DO--TH 20/05\\
SAGEX--20--09\\
June 2020 \\
\vspace{1.0cm}

\begin{center}
  {\bf \Large 
    Heavy-flavor PDF evolution and variable-flavor number scheme uncertainties in deep-inelastic scattering \\
  }

  \vspace{1.25cm}
 {\large
   S.~Alekhin$^{\, a,b}$,
   J.~Bl\"umlein$^{\, c}$,
   and S.~Moch$^{\, a}$
   \\
 }
 \vspace{1.25cm}
 {\it
   $^a$ II. Institut f\"ur Theoretische Physik, Universit\"at Hamburg \\
   Luruper Chaussee 149, D--22761 Hamburg, Germany \\
   \vspace{0.2cm}
   $^b$Institute for High Energy Physics \\
   142281 Protvino, Moscow region, Russia\\
   \vspace{0.2cm}
   $^c$Deutsches Elektronensynchrotron DESY \\
   Platanenallee 6, D--15738 Zeuthen, Germany \\
 }
  \vspace{1.4cm}
  \large {\bf Abstract}
  \vspace{-0.2cm}
\end{center}
We consider a detailed account on the construction of the heavy-quark 
parton distribution functions for charm and bottom, starting from $n_f=3$ 
light flavors in the fixed-flavor number (FFN) scheme and by using the 
standard decoupling relations for heavy quarks in QCD. We also account for 
two-mass effects. Furthermore, different implementations of the variable-flavor-number 
(VFN) scheme in deep-inelastic scattering (DIS) are studied, with the particular 
focus on the resummation of large logarithms in $Q^2/m_h^2$, the ratio the virtuality 
of the exchanged gauge-boson $Q^2$ to the heavy-quark mass squared $m_h^2$.
A little impact of resummation effects if found in the kinematic range of 
the existing data on the DIS charm-quark production so that they can be 
described very well within the FFN scheme. Finally, we study the theoretical 
uncertainties associated to the VFN scheme, which manifest predominantly at 
small~$Q^2$.
\end{titlepage}

\section{Introduction}
\label{sec:intro}

The process of deep-inelastic scattering (DIS) of leptons off a nucleon target  
provides important information on the nucleon structure and the parton content.
Therefore, it plays a central role in the determination of the 
parton distribution functions (PDFs), especially for the proton PDFs~\cite{Accardi:2016ndt}. 
At large values of Bjorken $x$ the DIS data constrain the valence quark distributions, 
while at small $x$ they are sensitive to the sea-quark and gluon distributions. 
In addition, the DIS cross sections at small $x$ contain substantial contributions from charm and bottom quarks.
The virtuality $Q^2$ of the exchanged gauge-boson is the other important kinematic variable in DIS.
It offers a wide range of scales to probe, for instance, in electron-proton scattering 
the parton dynamics inside the proton.
Depending on the value of $Q^2$, different theoretical
descriptions of DIS within Quantum Chromodynamics (QCD) may be applied.
This concerns in particular the number $n_f$ of active quark flavors 
and the treatment of the heavy quarks, as charm and bottom.

At low scales, when $Q^2$ is of the order of the 
heavy quark mass squared $m_h^2$, one typically works with $n_f = 3$ massless quark flavors. 
Then, the proton structure function is composed only out of light-quark PDFs for up, down and strange and of the gluon PDF. 
Massive quarks appear in the final state only or contribute as purely virtual corrections. 
At higher scales, for $Q^2 \gg m_c^2, m_b^2$ compared to the charm and bottom quark 
masses squared, additional dynamical degrees of freedom lead to
theories with $n_f = 4$ or $5$ effectively light flavors, 
depending on whether charm is considered massless, or even both, charm and bottom.
The massive renormalization group equations rule these dynamics and provide the corresponding 
scale evolution, linking to $n_f = 3$ massless quarks at very low virtualities.
In general, the transition for the flavor dependence of the strong coupling $\alpha_s$,
i.e., $\alpha_s(n_f) \to \alpha_s(n_f+1)$, 
is achieved at some matching scale $\mu_0$ with the decoupling relations~\cite{Appelquist:1974tg} 
which can be implemented perturbatively in QCD.
These decoupling relations introduce a logarithmic dependence on the heavy quark masses $m_c$, $m_b$.
In a similar manner, this is realized for the PDF $f_i$ of a parton $i$
with the help of suitable heavy-quark operator matrix elements (OMEs)~\cite{Buza:1996wv,Bierenbaum:2009mv},
which, in the perturbative expansion, also depend logarithmically on the heavy-quark masses.
The transition $f_i(n_f) \to f_i(n_f+1)$, again at a matching scale $\mu_0$,  
implies also the introduction of new heavy-quark PDFs for charm or bottom when they become effectively light
flavors, and can then be considered as effective dynamical degrees of freedom inside the proton.

For a given fixed value of $n_f$, and having decoupled the heavy quarks in an appropriate manner, 
one may define the fixed-flavor number (FFN) scheme.
In the FFN scheme used in the ABMP16 global fit of proton PDFs~\cite{Alekhin:2017kpj}, 
only light quarks and gluons are considered in the initial state, while 
heavy quarks appear in the final state as a result of the hard scattering of the incoming massless partons.
Existing data on the heavy-quark DIS production are 
well described by the FFN scheme with $n_f = 3$, see Refs.~\cite{Accardi:2016ndt,H1:2018flt}. 
However, many PDF fits, like those of CT18~\cite{Hou:2019efy}, MMHT14~\cite{Harland-Lang:2014zoa} 
and NNPDF3.1~\cite{Ball:2017nwa} employ various different versions of the so-called variable-flavor-number (VFN) scheme. 
In the VFN scheme the quark flavors charm and bottom are considered also in
the initial state from a certain mass scale onward and are dealt with as partonic components in the proton. 
As a consequence, the original distributions $f_i(n_f)$ are mapped into the distributions $f_i(n_f+1)$ 
at a chosen scale $\mu_0$, cf.~\cite{Buza:1996wv}.
In addition, the VFN scheme effectively performs a resummation of logarithms
in the ratio $Q^2/m_c^2$ (or $Q^2/m_b^2$) through the parton evolution 
equations for the charm (or bottom) PDF~\cite{Shifman:1977yb}, although 
the corresponding logarithms are not necessarily large for realistic kinematics. 

The difference in modeling of the 
heavy quark contribution, i.e., the choice for the FFN or the VFN scheme, has an impact on the 
PDFs obtained in global fits~\cite{Accardi:2016ndt,Thorne:2014toa}. 
Therefore, a detailed comparison of the two approaches is mandatory in view of
the use of the respective PDF sets in QCD precision phenomenology. 

A particular prescription for a VFN scheme has been proposed in \cite{Buza:1996wv}, commonly referred to as BMSN-scheme.
In PDF fits, the VFN scheme using this approach yields results which are not very different from the ones in the FFN scheme~\cite{Alekhin:2009ni}.
This happens due to a smooth transition between the $n_f$- and $(n_f+1)$-flavor regimes
at the matching scales $\mu_0=m_c$ and $\mu_0=m_b$, respectively, which is imposed in the BMSN ansatz. 
However, the BMSN prescription is based on heavy-quark PDFs, i.e., charm and bottom, which are 
derived with the help of fixed-order matching conditions. 
Therefore, the results of our previous study~\cite{Alekhin:2009ni} 
cannot be directly compared to the PDF fits in  
Refs.~\cite{Hou:2019efy,Harland-Lang:2014zoa,Ball:2017nwa} which apply heavy-quark PDF evolution.

In the present article we study the phenomenology of 
a modified BMSN prescription, which also includes the scale evolution of heavy-quark PDFs
in order to clarify the basic features of such VFN schemes.
Our studies are limited to the case of DIS charm-quark production, 
since this process is most essential phenomenologically and, at the same time, 
a representative case.

The paper is organized as follows. 
Basic features of QCD factorization, the VFN scheme and the BMSN prescription are outlined in Sec.~\ref{sec:form}.
In Sec.~\ref{sec:evol} we describe the particularities introduced by the heavy-quark PDF evolution and 
Sec.~\ref{sec:pheno} contains the benchmarking of various factorization schemes 
based on existing data for DIS charm-quark production. 
We address implications of VFN schemes for predictions at hadron colliders 
Sec.~\ref{sec:colliders} and conclude in Sec.~\ref{sec:concl}. 
Technical details of the various implementations of heavy-quark schemes are 
summarized in App.~\ref{sec:appA}.

\section{Heavy-quark PDFs}
\label{sec:form}

The dynamics of massless partons in the proton are parameterized in terms of the PDFs $f_i$ 
with $i=u,d,s,g$ for up, down, strange quarks and the gluon. 
These define the set of flavor-singlet quark and gluon PDFs, $q^{s}$ and $g$, 
\begin{eqnarray}
\label{eq:qs+g-PDFs}
q^{s}(n_f, \mu^2) \,=\, \sum_{l=1}^{n_f} (f_l(n_f, \mu^2) + \bar{f}_l(n_f), \mu^2)
\, ,
\qquad\qquad
g(n_f, \mu^2) \,=\, f_g(n_f, \mu^2)
\, ,
\end{eqnarray}
where $\mu$ denotes the factorization scale and 
we suppress the dependence on the momentum fractions $x$ here and below.

Using standard QCD factorization~\footnote{A variant of the VFN scheme is used 
in the NNPDF3.1 fit of Ref.~\cite{Ball:2017nwa}, where the heavy-quark PDFs are parameterized 
by some functional form, which is then fitted to the data.}, 
the PDFs for the heavy quarks charm and bottom ($h =c, b$) at the scale $\mu$ 
in the \msbar\ scheme and using on-shell renormalization for the mass $m_h$ 
are then constructed 
from the quark-singlet and gluon PDFs in Eq.~(\ref{eq:qs+g-PDFs}) 
and the heavy-quark OMEs $A_{ij}$ as follows~\cite{Buza:1996wv,Bierenbaum:2009mv}
\begin{eqnarray}
  \label{eq:VFNS-hq}
  f_{h+\bar h}(n_f+1, \mu^2)
  &=& {A_{hq}^{ps}\Big(n_f, \frac{\mu^2}{m_h^2}\Big)} \otimes {q^{s}(n_f, \mu^2)}
  + {A_{hg}^{s}\Big(n_f, \frac{\mu^2}{m_h^2}\Big)} \otimes {g(n_f, \mu^2)}
  \, ,
\end{eqnarray}
where $h=c,b$ and `$\otimes$' denotes the Mellin convolution in the momentum fractions $x$. 
Typically, the matching conditions are imposed at the scale $\mu_0 = m_h$, 
and we further assume that $f_{h + {\bar h}} = 0$  at scales $\mu \le m_h$.
In addition, the transition $\{q^{s}(n_f), g(n_f)\} \to \{q^{s}(n_f+1), g(n_f+1)\}$ 
for the set of the light-quark singlet and the gluon distributions 
with the respective heavy-quark OMEs has to account for 
operator mixing in the singlet sector
\begin{eqnarray}
  \label{eq:VFNS-lq}
  q^{s}(n_f+1,\mu^2)
  &=& \left[ A_{qq,h}^{ns}\Bigl(n_f, \frac{\mu^2}{m_h^2}\Bigr) + 
    A_{qq,h}^{ps} \Bigl(n_f, \frac{\mu^2}{m_h^2}\Bigr) +
    A_{hq}^{ps} \Bigl(n_f, \frac{\mu^2}{m_h^2}\Bigr) \right] \otimes q^{s}(n_f,\mu^2)
 \nonumber\\ &&
  +\left[A^{s}_{qg,h}\Bigl(n_f, \frac{\mu^2}{m_h^2}\Bigr)
    + A^{s}_{hg}\Bigl(n_f, \frac{\mu^2}{m_h^2}\Bigr) \right] \otimes g(n_f,\mu^2)
  \, ,
  \\
  \label{eq:VFNS-g}
  g(n_f+1, \mu^2)
  &=& A_{gq,h}^{s}\Bigl(n_f,\frac{\mu^2}{m_h^2}\Bigr) \otimes q^{s}(n_f,\mu^2)
  + A_{gg,h}^{s}\Bigl(n_f,\frac{\mu^2}{m_h^2}\Bigr) \otimes g(n_f,\mu^2)
  \, ,
\end{eqnarray}
with $h=c,b$, see Refs.~\cite{Buza:1996wv,Bierenbaum:2009mv}, 
also for matching relations for the non-singlet distributions.

The perturbative expansion of the OMEs in powers of the strong coupling constant $\alpha_s$ 
reads (using $a_s = \alpha_s/(4\pi)$ as a short-hand),
\begin{eqnarray}
  \label{eq:OMEexp}
  A_{ij}
  \; = \;
  \delta_{ij} +
  \sum\limits_{k=1}^{\infty}\, 
  a_s^k \,
  A_{ij}^{(k)}
  \; = \;
  \delta_{ij} +
  \sum\limits_{k=1}^{\infty}\, 
  a_s^k \,
  \sum\limits_{\ell=0}^{k}\, a^{(k,\ell)}_{ij}\, 
  \ln^{\,\ell}\left(\frac{\mu^2}{m_h^2}\right)
  \, ,
  \qquad
\end{eqnarray}
where the expressions $a^{(k,0)}_{ij}$ contain the information, which is genuinely new at the $k$-th order.
The leading-order (LO) and next-to-leading order (NLO) contributions to the OMEs 
are given by the coefficients at order $a_s$ and $a_s^2$ in Eq.~(\ref{eq:OMEexp}), respectively.
They have been determined analytically in closed form 
in Refs.~\cite{Buza:1995ie,Buza:1996wv,Bierenbaum:2007qe,Bierenbaum:2009zt}~\footnote{
  The initial calculation of the two-loop OMEs $A_{hg}^{s,\, (2)}$ and
  $A_{gg,h}^{s,\, (2)}$ in Ref.~\cite{Buza:1996wv} was incomplete,
  cf. Ref.~\cite{Bierenbaum:2009zt}.}. 
At next-to-next-to-leading (NNLO) the heavy-quark OMEs are known
either exactly or to a good approximation~\cite{Bierenbaum:2009mv,Ablinger:2010ty,Kawamura:2012cr,Ablinger:2014lka,Ablinger:2014nga,Alekhin:2017kpj}.
This includes specifically the non-singlet and pure-singlet constant parts $a_{hq}^{(3,0)}$ in Eq.~(\ref{eq:OMEexp}) 
and the term $a_{hg}^{(3,0)}$ at order $a_s^3$. 
In the latter case an approximation based on fixed Mellin moments~\cite{Bierenbaum:2009mv} 
with a residual uncertainty in the small-$x$ region has been given in Ref.~\cite{Alekhin:2017kpj,Kawamura:2012cr}.

It should be noted that, the decoupling relations in Eqs.~(\ref{eq:VFNS-hq})--(\ref{eq:VFNS-g}) 
assume the presence of a single heavy quark at each step only.
Thus, the bottom-quark contributions are ignored in the transition from $n_f=3$ to $4$ 
and in the construction of the charm-quark PDF.
However, starting at two-loop order, the perturbative corrections to the heavy-quark OMEs contain 
graphs with both, charm- and bottom-quark lines.  
With the ratio of masses $(m_c/m_b)^2 \approx 1/10$, 
charm quarks generally cannot be taken as massless at the scale of the bottom-quark. 
Such two-mass contributions to the heavy-quark OMEs have been computed 
recently~\cite{Ablinger:2017err,Ablinger:2017xml,Ablinger:2018brx}. 
At three--loop order (and beyond), these corrections can neither be attributed to the charm- nor to the bottom-quark PDFs separately.
Rather, one has to decouple charm and bottom quarks together at some large scale and the corresponding 
VFN scheme, i.e. the simultaneous transition with two massive quarks, $f_i(n_f) \to f_i(n_f+2)$, 
has been discussed recently in Ref.~\cite{Blumlein:2018jfm}.
This proceeds in close analogy to the simultaneous decoupling of bottom and charm
quarks in the strong coupling constant $\alpha_s$, 
see for instance Ref.~\cite{Grozin:2011nk}. 
We will elaborate on these aspects further below.

First, we will limit our studies to the case of the charm-quark PDF 
and apply Eqs.~(\ref{eq:VFNS-hq})--(\ref{eq:VFNS-g}) to change from $n_f=3$ to $4$.
At LO only the heavy-quark OME $A_{hg}^{s}$ contributes 
and the coefficients are
\begin{equation}
  \label{eq:Ahg-one-loop}
  a_{hg}^{(1,0)}(x) \,=\, 0
  \, ,
  \qquad
  \qquad
  a_{hg}^{(1,1)}(x) \,=\, 4 T_f (1 - 2 x + 2 x^2)
  \,=\, \frac{P^{(0)}_{qg}(x)}{n_f}\,
  \, , 
\end{equation}
i.e., the constant term of the unrenormalized massive OME $A_{hg}^{s,\, (1)}$ vanishes and the logarithmic one 
with $T_f = 1/2$ is proportional to the LO quark-gluon splitting function $P^{(0)}_{qg}$ 
in the normalization of Ref.~\cite{Vogt:2004mw}. 
For four active flavors we abbreviate the charm PDF in Eq.~(\ref{eq:VFNS-hq})
as $c(x,\mu^2) \,\equiv\, f_{c+\bar c}(4,x,\mu^2)$ and consider its perturbative expansion
%
\begin{equation}
\label{eq:charm-pdf-def}
c(x,\mu^2) \,=\, c^{(1)}(x,\mu^2) + c^{(2)}(x,\mu^2) + \dots,
\end{equation}
where the LO term $c^{(1)}(x,\mu^2)$ has a particularly simple form,
\begin{equation}
\label{eq:cqlo}
c^{(1)}(x,\mu^2) \,=\,
a_s(\mu^2)\, \ln\left(\frac{\mu^2}{m_c^2}\right)\, 
\int_x^1 \frac{dz}{z}\,\, a_{hg}^{(1,1)}(z)\,\, g\left(n_f=3,\frac{x}{z},\mu^2\right)
\, .
\end{equation}
%
Here, $g$ denotes the gluon PDF in the 3-flavor scheme.
This expression is used in the BMSN prescription~\cite{Buza:1996wv} of the VFN scheme 
and determines the charm-quark distribution at all scales $\mu \geq m_c$ 
at fixed-order perturbation theory (FOPT).

On the contrary, other VFN prescriptions, like 
ACOT~\cite{Aivazis:1993pi,Kramer:2000hn,Tung:2001mv}, 
FONLL~\cite{Forte:2010ta} or RT~\cite{Thorne:1997ga} 
use Eq.~(\ref{eq:cqlo}) as a boundary condition for $c(x,\mu^2)$ 
at $\mu=m_c$ and derive the scale dependence with the help 
of the standard QCD evolution equations (DGLAP) for massless quarks.
The evolution resums logarithmic terms to all orders, 
so that the charm-quark distribution acquires additional higher order contributions 
which are not present in the FOPT expression in Eq.~(\ref{eq:cqlo}).
In order to illustrate the numerical difference between 
these two approaches, we consider the derivative of $c(x,\mu^2)$
%
\begin{eqnarray}
  \label{eq:cloder}
  \frac{dc^{(1)}(x,\mu^2)}{d\ln\mu^2} 
  &=& 
  a_s(\mu^2)\int_x^1 \frac{dz}{z}\,\, a_{hg}^{(1,1)}(z)\,\,g\left(\frac{x}{z},\mu^2\right)
  \,+\, \left(\frac{da_s}{d\ln\mu^2}\right) \frac{c^{(1)}(x,\mu^2)}{a_s} 
  \nonumber \\
  & &
  \,+\, a_s(\mu^2)\ln\left(\frac{\mu^2}{m_c^2}\right) \int_x^1 \frac{dz}{z} \,\,a_{hg}^{(1,1)}(z)\,\, \dot{g}\left(\frac{x}{z},\mu^2\right)
  \, ,
\end{eqnarray}
%
where 
$\dot{g}(x,\mu^2)\equiv dg(x,\mu^2)/d\ln\mu^2$.

%
\begin{figure}
  \centering
  \includegraphics[width=\textwidth,height=0.5\textwidth]{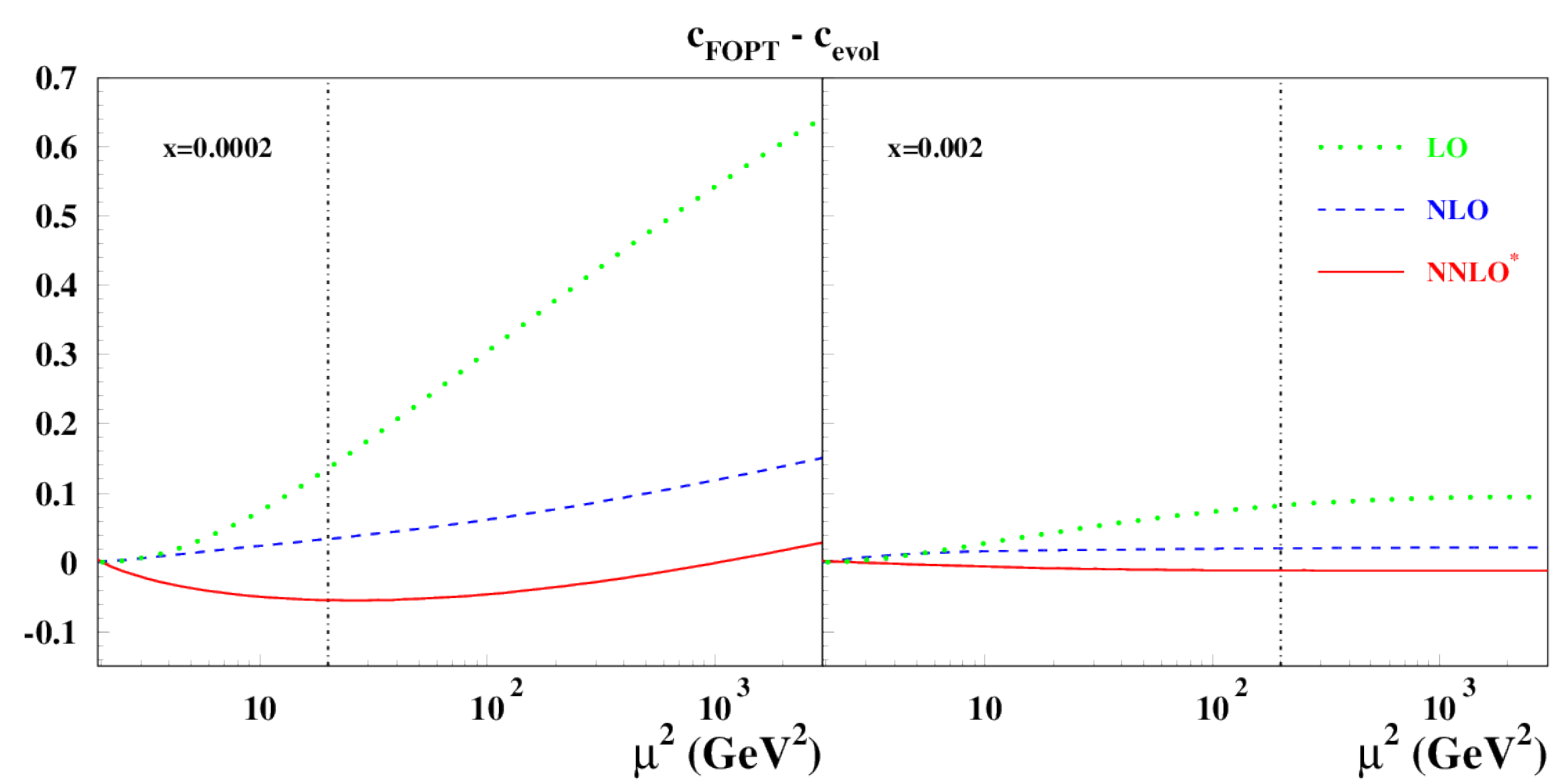}
  \caption{\small 
The difference between the evolved $c$-quark distributions and the ones obtained with the FOPT conditions 
in various orders of QCD (LO: dots, NLO: dashes, and NNLO$^{\ast}$: solid lines) 
versus the factorization scale $\mu$ and at representative values 
of the parton momentum fraction $x$ (left: $x=0.0002$, right: $x=0.002$)
taking the matching scale $\mu_0=m_c=1.4~{\rm GeV}$, 
where  $m_c$ is the pole mass of $c$-quark mass.
The vertical dash-dotted lines display the upper margin for the HERA collider kinematics.
}
    \label{fig:pdfevol}
\end{figure}

\begin{figure}[t!]
\centering 
\includegraphics[width=1.0\textwidth]{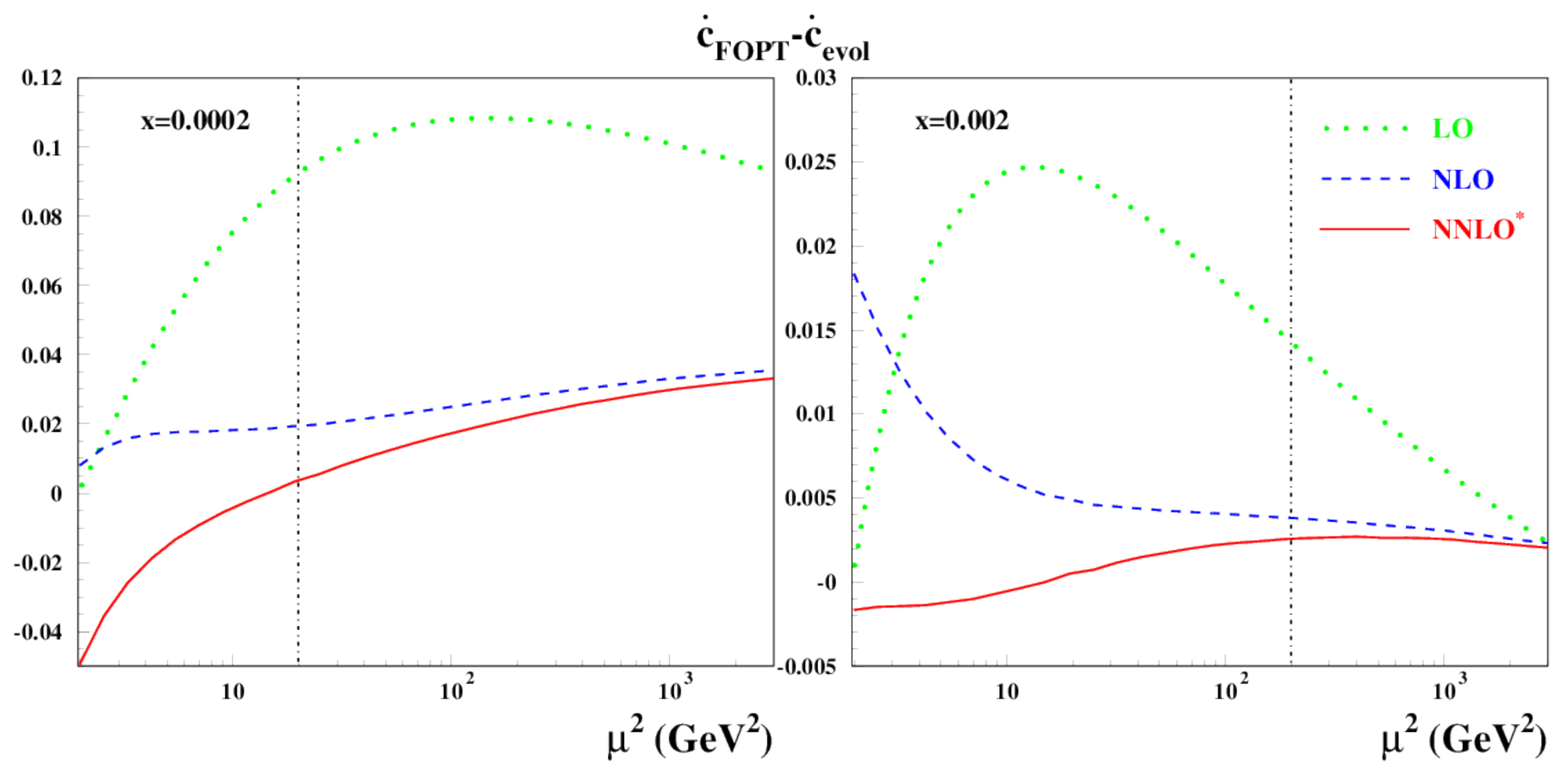}
\hfill
\caption{\label{fig:pdfder} 
The same as in Fig.~\ref{fig:pdfevol} for the scale derivatives of the charm-quark PDF,  
$\dot{c}(x,\mu^2) \equiv dc(x,\mu^2)/d\ln\mu^2$.}
\end{figure}

The first term in Eq.~(\ref{eq:cloder}) corresponds to the right hand side of
the standard DGLAP evolution equations, recall Eq.~(\ref{eq:Ahg-one-loop}),
i.e., $a_{hg}^{(1,1)}$ is proportional to $P^{(0)}_{qg}$.
The second and the third term, however, account for the difference 
between the FOPT distributions and the evolved ones. 
These terms vanish at the matching scale $\mu_0=m_c$ as they should by definition. 
For scales $\mu > m_c$ the second term proportional to the 
QCD $\beta$ function is negative, since $da_s/d\ln \mu^2 = \beta(a_s)/(4\pi) < 0$.
However, the net effect of the difference between the FOPT and the DGLAP evolved distributions
shown in Fig.~\ref{fig:pdfevol} on the left is positive at small $x$ and driven by $\dot{g}$ in the third term.
Only at large $x$, where the gluon PDF is negligible, the 
term proportional to $\beta(a_s)$ dominates and the net difference 
between the FOPT and the DGLAP evolved distributions is negative.

The matching conditions for the charm-quark at NLO are more involved.
The NLO term $c^{(2)}(x,\mu^2)$ in Eq.~(\ref{eq:charm-pdf-def}) has the form
%
\begin{eqnarray}
\label{eq:cq-two-loop}
c^{(2)}(x,\mu^2) &=&
a_s^2(\mu^2) \int_x^1 \frac{dz}{z}\,\, A_{hg}^{s,\, (2)}\left(n_f=3,z,\frac{\mu^2}{m_c^2}\right)\,\, g\left(n_f=3,\frac{x}{z},\mu^2\right)
  \nonumber \\
  & & 
  \,+\, 
a_s^2(\mu^2) \int_x^1 \frac{dz}{z}\,\, A_{hq}^{ps,\, (2)}\left(n_f=3,z,\frac{\mu^2}{m_c^2}\right)\,\, q^{s}\left(n_f=3,\frac{x}{z},\mu^2\right)
\, .
\end{eqnarray}
%
It includes the NLO corrections to the massive OMEs $A_{hg}^{s,\, (2)}$ and
$A_{hq}^{ps,\, (2)}$ for $n_f=3$, see Eq.~(\ref{eq:OMEexp}),
the gluon and the quark-singlet PDFs, $g$ and $q^{(s)}$, 
are taken in the 3-flavor scheme again, cf. Eq.~(\ref{eq:qs+g-PDFs}).
Since $a_{hg}^{(2,0)}$ and $a_{hq}^{(2,0)}$ in Eq.~(\ref{eq:OMEexp}) are
non-zero in the \msbar\ scheme, $c(x,\mu^2)$ at NLO does not vanish anymore
at the matching scale $\mu_0=m_c$, see the off-set in Fig.~\ref{fig:pdfevol} on the right.
%
\begin{figure}[t!]
\centering 
\includegraphics[width=1.0\textwidth]{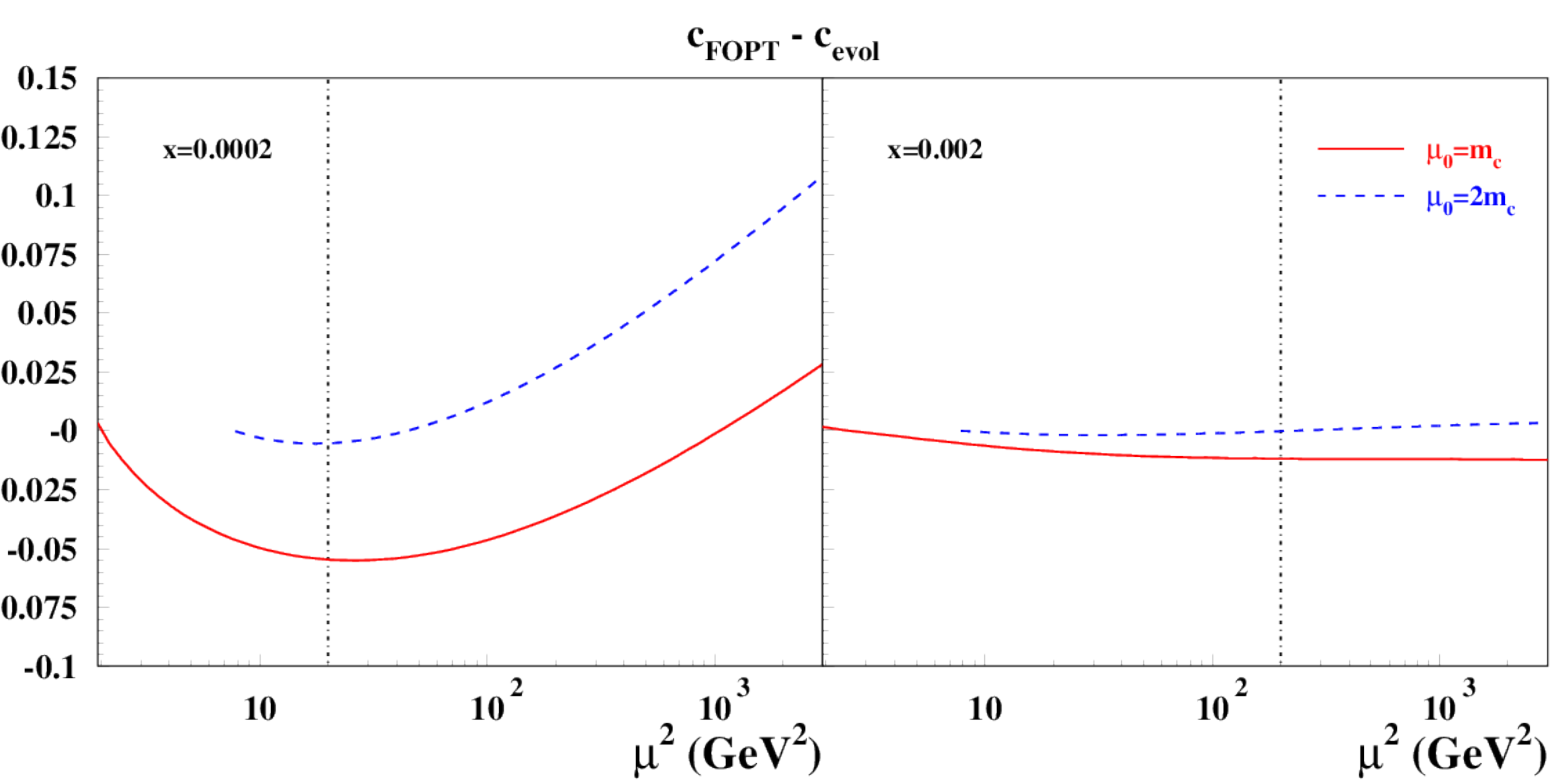}
\hfill
\caption{\label{fig:pdfevolth} 
The same as in Fig.~\ref{fig:pdfevol} at NNLO$^{\ast}$ and different 
values of the matching scale $\mu_0$ (solid line: $\mu_0=m_c$, dashes: $\mu_0=2m_c$).
}
\end{figure}
%

The comparison of the charm-quark FOPT distributions at NLO based on 
Eqs.~(\ref{eq:cqlo}) and (\ref{eq:cq-two-loop}) 
and the evolved ones, using $c(x,\mu^2)$ only as the boundary condition at the matching scale, 
shows in Fig.~\ref{fig:pdfevol} qualitatively the same pattern as at LO,
although the numerical differences are smaller now.
At small $x$, driven by the scale derivative $\dot{g}$ of the gluon PDF, the
FOPT distributions are larger while at large $x$ the terms proportional to 
$\beta(a_s)$ dominate and the DGLAP evolved distributions are larger.
These observations can be expressed in quantitative form 
through the scale derivative of the NLO term $c^{(2)}(x,\mu^2)$, which reads
%
\begin{eqnarray}
  \label{eq:cnloder}
  \frac{dc^{(2)}(x,\mu^2)}{d\ln\mu^2} 
  &=& 
  a_s^2(\mu^2)
  \int_x^1 \frac{dz}{z}\,\, \left(a_{hg}^{(2,1)}(z) + 2 \ln\left(\frac{\mu^2}{m_c^2}\right) a_{hg}^{(2,2)}(z)\right)\,\, g\left(\frac{x}{z},\mu^2\right)
  \nonumber \\
  & &
  \,+\, 
  a_s^2(\mu^2)
  \int_x^1 \frac{dz}{z}\,\, \left(a_{hq}^{(2,1)}(z) + 2 \ln\left(\frac{\mu^2}{m_c^2}\right) a_{hq}^{(2,2)}(z)\right)\,\, q^{s}\left(\frac{x}{z},\mu^2\right)
  \,+\, 
  2 \left(\frac{da_s}{d\ln\mu^2}\right) \frac{c^{(2)}(x,\mu^2)}{a_s} 
  \nonumber \\
  & &
  \,+\, 
  a_s^2(\mu^2)
  \int_x^1 \frac{dz}{z}\,\, A_{hg}^{s,\, (2)}\left(z,\frac{\mu^2}{m_c^2}\right)\,\, \dot{g}\left(\frac{x}{z},\mu^2\right)
  + 
  a_s^2(\mu^2)
  \int_x^1 \frac{dz}{z}\,\, A_{hq}^{ps,\, (2)}\left(z,\frac{\mu^2}{m_c^2}\right)\,\, \dot{q}^{s}\left(\frac{x}{z},\mu^2\right)
  \, ,
  \nonumber \\
\end{eqnarray}
%
where, again, 
$\dot{g}(x,\mu^2)\equiv dg(x,\mu^2)/d\ln\mu^2$ and $\dot{q}^{s}(x,\mu^2)\equiv dq^{s}(x,\mu^2)/d\ln\mu^2$.

Here, the first two terms in the right hand side contain the expressions 
used in the standard DGLAP equations to evolve the charm-quark PDF, 
since the NLO splitting functions $P^{(1)}_{qg}$ and $P^{(1)}_{qq}$ appear 
in the terms $a_{hg}^{(2,1)}$ and $a_{hq}^{(2,1)}$, cf.~\cite{Buza:1996wv,Bierenbaum:2009mv}.
However, there are also other contributions, since the heavy-quark OMEs enjoy
their own (massive) renormalization group equation.
In addition, the full expression $dc(x,\mu^2)/d\ln\mu^2$ at NLO contains, of course, also 
the terms from $\dot{c}^{(1)}(x,\mu^2)$ in Eq.~(\ref{eq:cloder})
expanded to higher order in $a_s$, for example the term proportional to $\beta(a_s)$.
In summary, these terms are responsible for decreasing the difference
between the FOPT and the evolved distributions at NLO in Fig.~\ref{fig:pdfevol}.

As a further variant in the study of the DGLAP evolved charm-quark PDF, one
can perform the evolution using the full NNLO splitting functions $P^{(2)}_{ij}$ 
of Ref.~\cite{Vogt:2004mw} starting at the matching scale $\mu_0=m_c$ 
from the boundary condition for $c(x,m_c^2)$ at NLO in Eqs.~(\ref{eq:cqlo}) and (\ref{eq:cq-two-loop}).
We denote this variant as NNLO$^{\ast}$, since there is 
a mismatch in the orders of perturbation theory between the heavy-quark OMEs
and the accuracy of the evolution equations.
The difference with the NLO variant is due to terms which are formally 
of higher order, but nevertheless have significant numerical impact 
at small $x$ as shown in Fig.~\ref{fig:pdfevol}.
There, the FOPT distributions at NLO and the evolved ones at NNLO$^{\ast}$
accuracy are very similar in the entire $\mu$-range.
Only at large $x$, the increased order in the DGLAP evolution is negligible.

In Fig.~\ref{fig:pdfder} we display the scale derivatives of the charm-quark
PDF $\dot{c}(x,\mu^2) \equiv dc(x,\mu^2)/d\ln\mu^2$ calculated using 
Eqs.~(\ref{eq:cloder}) and (\ref{eq:cnloder}).
We consider the difference of $\dot{c}(x,\mu^2)$ determined in FOPT, $\dot{c}_{\rm{FOPT}}$,
and the one evolved with the standard DGLAP evolution, $\dot{c}_{\rm{evol}}$, 
choosing $n_f=4$ and starting from the expressions in Eqs.~(\ref{eq:cqlo}) and (\ref{eq:cq-two-loop}) 
at the matching scale $\mu_0=m_c=1.4~{\rm GeV}$.
Evidently, at LO the difference $\dot{c}_{\rm{FOPT}}-\dot{c}_{\rm{evol}}$ has to vanish at
the matching scale, while 
at NLO or in the NNLO$^{\ast}$ variant some finite off-set at
$\mu_0=m_c=1.4~{\rm GeV}$ remains.
Remarkably, the results at NLO and at NNLO$^{\ast}$, i.e., using NLO boundary
conditions from Eqs.~(\ref{eq:cqlo}) and (\ref{eq:cq-two-loop}) and NNLO
splitting functions in the evolution of $\dot{c}_{\rm{evol}}$ are very 
different at low factorization scales and only converge above 
$\mu^2 \gtrsim 10^2 \dots 10^3$~GeV$^2$, depending on the value of $x$.
These large scales, however, at which the NLO and the NNLO$^{\ast}$ variants 
become of similar size, are typically well outside the kinematic range of the
HERA collider, whose upper limit is indicated by the vertical arrow.
These findings indicate that there is a substantial numerical uncertainty 
in the VFN prescriptions 
ACOT~\cite{Aivazis:1993pi,Kramer:2000hn,Tung:2001mv}, 
FONLL~\cite{Forte:2010ta} or RT~\cite{Thorne:1997ga} 
due to the order of the QCD evolution applied.
In particular the additional higher order terms in the NNLO$^{\ast}$ variant
do have a sizable effect within the $\mu$-range covered by experimental data
on DIS charm-quark production and, hence, on the quality of the description of
those data in a fit using such VFN prescriptions.

An additional source of uncertainty in the VFN scheme concerns the choice of 
the matching scale $\mu_0$. Conventionally it is set to the 
corresponding heavy-quark mass, $m_c$ and $m_b$ for the 4- and 5-flavor PDFs, respectively. 
A variation of $\mu_0$ leads to a modification of the shape of the evolved heavy-quark PDFs, 
whereas, in contrast, the FOPT ones remain unchanged by construction. 
Therefore, for $\mu_0 > m_h$ a difference between the FOPT and the evolved 
heavy-quark PDFs is generally becoming smaller, in particular within 
the phase-space region covered by existing data, cf. Fig.~\ref{fig:pdfevolth}. 
Such a variation of $\mu_0$ also implies the use of the FFN scheme 
to describe data in a wider kinematic range,
e.g., up to $\mu_0 = 2m_c$ instead of $\mu_0 = m_c$ for the illustration in Fig.~\ref{fig:pdfevolth}.
Therefore, the uncertainty due to a variation of matching scale is not completely independent 
from those related to the choice of heavy-quark PDFs employed in the VFN scheme.
%
\begin{figure}[t!]
\centering 
\includegraphics[width=1.0\textwidth]{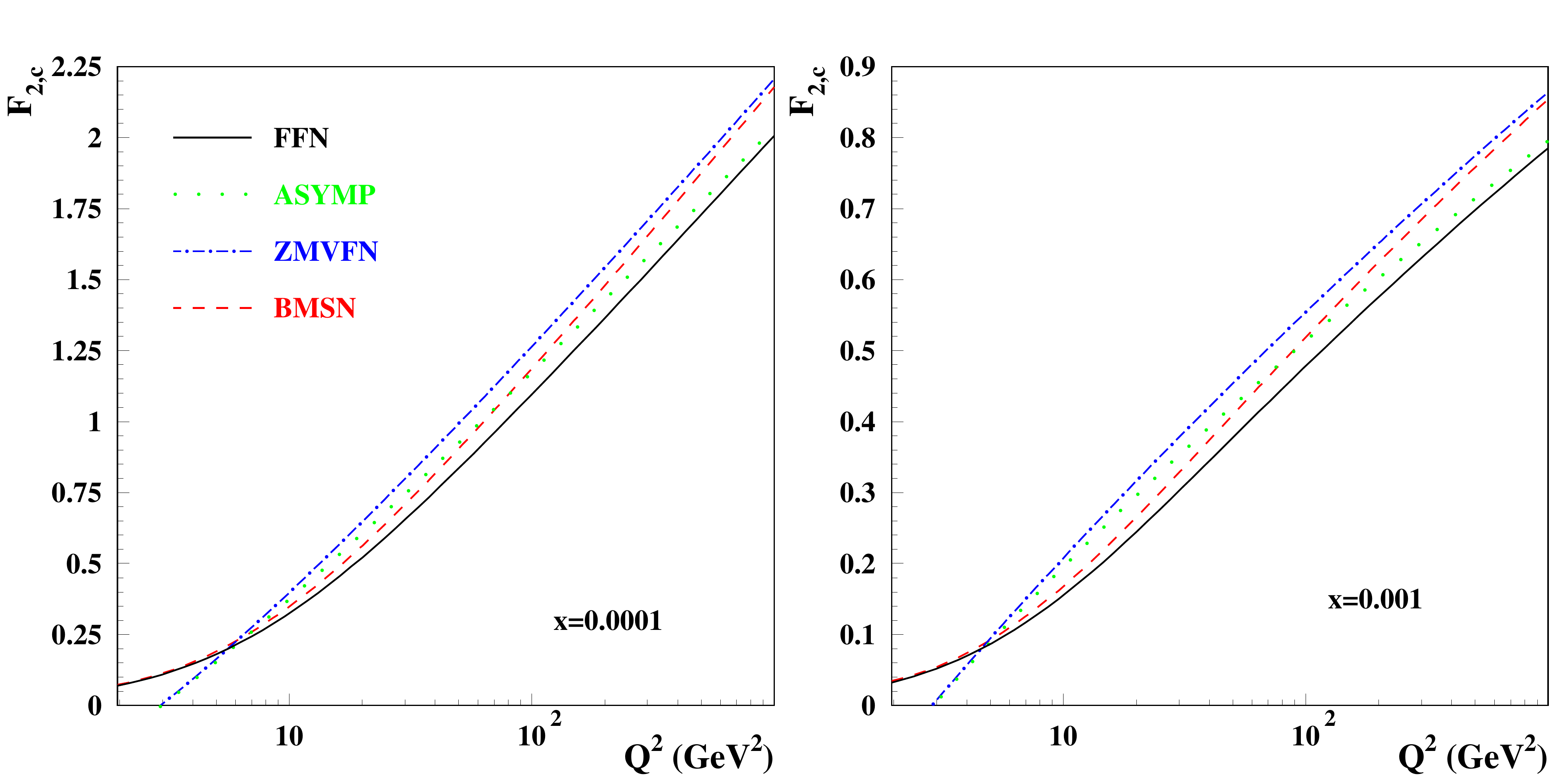}
\hfill
\caption{\label{fig:bmsn} 
The structure function $F_2^c$ for DIS $c$-quark production 
at values $x=0.0001$ (left) and $x=0.001$ (right) of the Bjorken variable 
versus momentum transfer $Q^2$ computed 
in the FFN scheme (solid lines), 
with the asymptotic expression of the FFN scheme (dots), 
the ZMVFN scheme (dash-dotted lines) and the BMSN prescription of the VFN scheme (dashes) and, 
using the FOPT $c$-quark distribution. 
}
\end{figure}
%
\begin{figure}[t!]
\centering 
\includegraphics[width=1.0\textwidth]{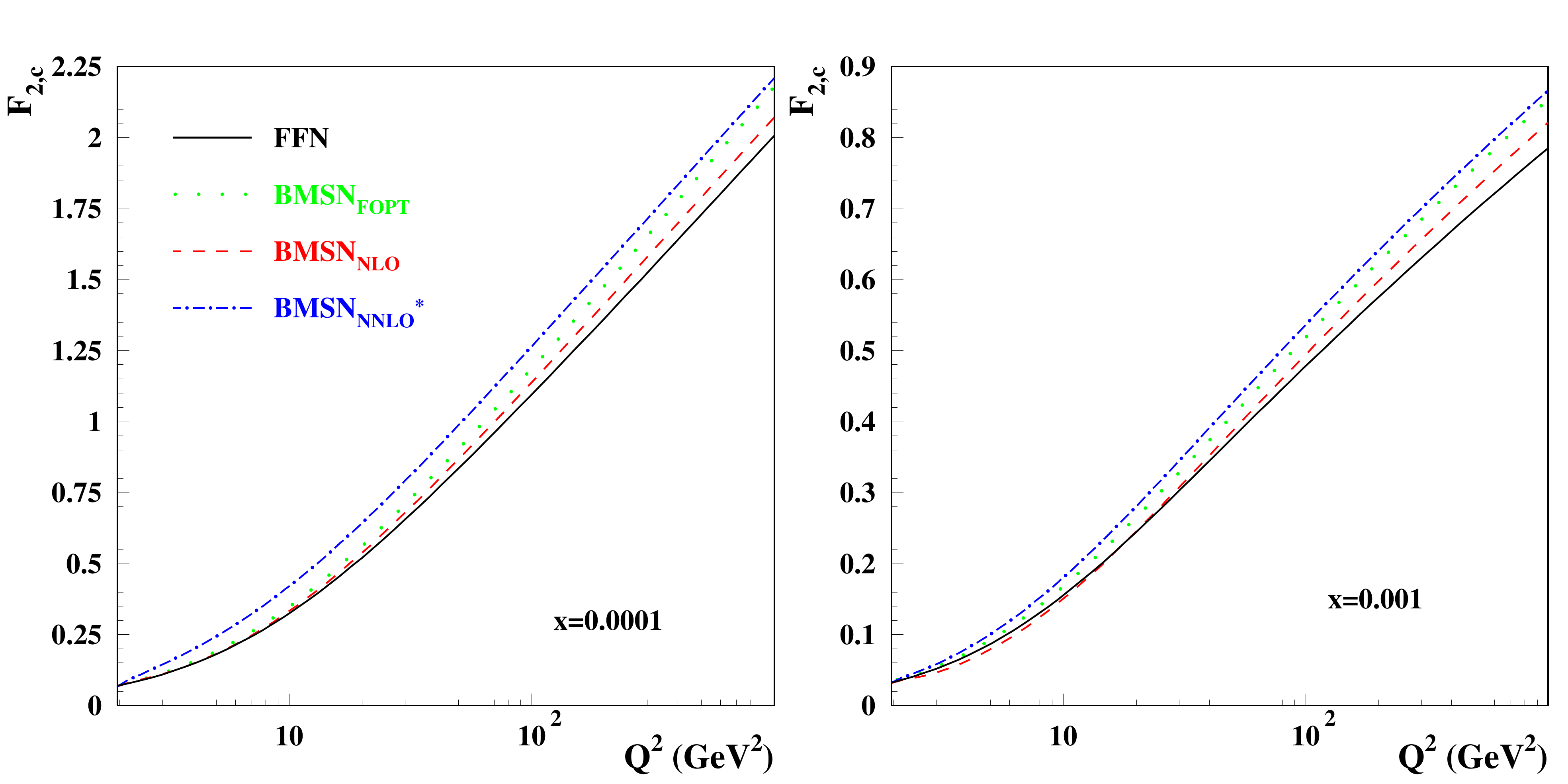}
\hfill
\caption{\label{fig:f2c} 
The same as in Fig.~\ref{fig:bmsn} computed using the BMSN prescription 
of the VFN scheme and various approaches for the generation of the $c$-quark 
distributions (FOPT: dots, NLO evolved: dashes, NNLO$^{\ast}$ evolved: 
dash-dotted lines) in comparison with the FFN scheme results (solid lines). 
}
\end{figure}
%
\begin{figure}[t!]
\centering 
\includegraphics[width=1.0\textwidth]{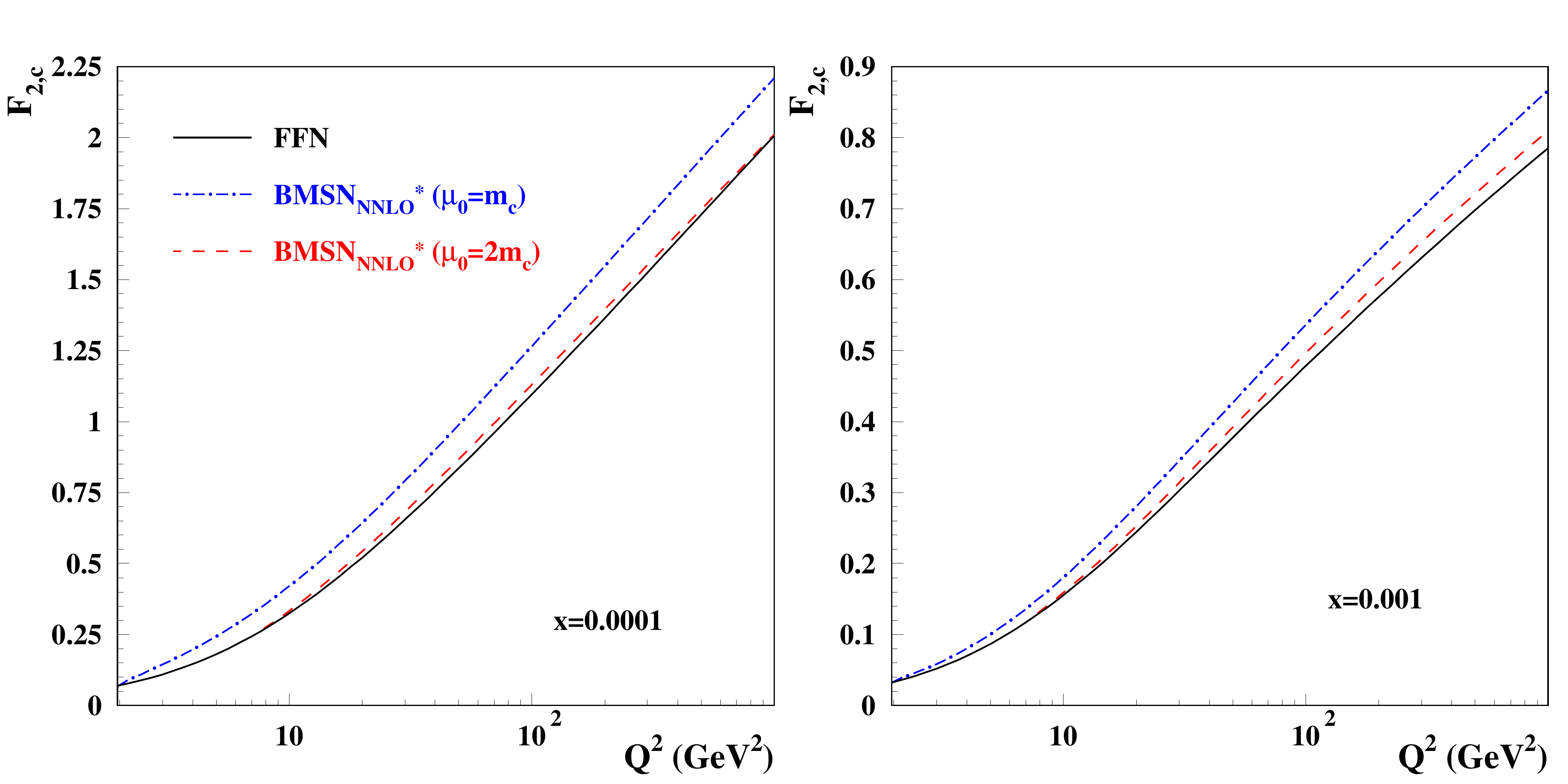}
\hfill
\caption{\label{fig:f2cth} 
The same as in Fig.~\ref{fig:f2c} for the NNLO$^{\ast}$ $c$-quark distributions
obtained with the matching scales $\mu_0=m_c$ (dash-dotted lines) and $\mu_0=2m_c$ (dashes).
}
\end{figure}
%

\section{BMSN prescription of the VFN scheme }
\label{sec:evol}

The heavy-quark distribution derived using the matching 
conditions Eqs.~(\ref{eq:VFNS-hq}), (\ref{eq:VFNS-lq}) enter the 
zero-mass VFN scheme (ZMVFN) expression for $F_{2,h}$
\begin{equation}  
\label{eq:zmvfn}
F_{2,h}^{ZMVFN} \,=\, 
\sum\limits _{k=0}^{\infty} a_s^k(n_f+1) \sum\limits_{i=q,g,h} C^{(k)}_{2,i}(n_f+1) \otimes f_i(n_f+1)
\, ,
\end{equation}  
where $C_{2,i}^{(k)}$ are the massless DIS Wilson coefficients at the $k$-th
order, which are known to next-to-next-to-next-to-leading order (N$^3$LO)~\cite{Vermaseren:2005qc}. 
This expression is valid at asymptotically large momentum transfer $Q^2 \gg m_h^2$, 
while it is unsuitable for scales $Q^2 \simeq m_h^2$ since the heavy-quark decoupling is not applicable.
Therefore, a realistic implementation of the VFN scheme commonly includes a combination 
of the ZMVFN expression in Eq.~(\ref{eq:zmvfn}) with the FFN one
\begin{equation}  
\label{eq:ffn}
F_{2,h}^{FFN} \,=\, 
\sum\limits _{k=1}^{\infty} a_s^k(n_f) \sum\limits _{i=q,g} H^{(k)}_{2,i}(n_f) \otimes f_i(n_f)
\, ,
\end{equation}  
where $H_{2,i}^{(k)}$ are the Wilson coefficients for the DIS heavy-quark production, 
all known exactly at NLO~\cite{Laenen:1992zk} and $H^{(3)}_{2,g}$ 
to a good approximation at NNLO~\cite{Kawamura:2012cr,Alekhin:2017kpj}.
Furthermore, in order to avoid double counting, a subtraction has to be carried out 
when combining Eqs.~(\ref{eq:zmvfn}) and (\ref{eq:ffn}).
For the BMSN prescription of the VFN scheme~\cite{Buza:1996wv} 
this subtraction arises from the asymptotic FFN expression as follows
\begin{equation}  
\label{eq:asymp}
F_{2,h}^{asy} \,=\,
\sum\limits _{k=1}^{\infty} a_s^k(n_f) \sum\limits _{i=q,g} H^{(k),asy}_{2,i}(n_f) \otimes f_i(n_f)
\, ,
\end{equation}  
where $H^{(k),asy}_{2,i}$ is derived from $H^{(k)}_{2,i}$ taken in the limit of $Q^2 \gg m_h^2$.
In summary BMSN prescription then reads
\begin{equation}
\label{eq:bmsn}
F_{2,h}^{BMSN} \,=\, F_{2,h}^{FFN}+F_{2,h}^{ZMVFN}-F_{2,h}^{asy}\, ,
\end{equation}  
where a factorization scale $\mu_F=m_h$ is used throughout.

The asymptotic Wilson coefficients $H^{asy}_{2,i}$ can be expanded into 
a linear combination of the massless Wilson coefficients $C_{2,i}$ and 
the massive OMEs~\cite{Buza:1995ie,Buza:1996wv,Bierenbaum:2009mv}.
For this reason, the asymptotic expression of Eq.~(\ref{eq:asymp}) 
coincides with the ZMVFN one of Eq.~(\ref{eq:zmvfn}), 
when the FOPT matching conditions Eqs.~(\ref{eq:VFNS-hq})--(\ref{eq:VFNS-g}) are employed, 
up to the subleading non-singlet terms and the difference between 
$a_s^k(n_f+1)$ and $a_s^k(n_f)$~\cite{Alekhin:2009ni}.
The latter exhibits a small discontinuity at $Q^2 \simeq m_h^2$ 
beyond one loop~\cite{Schroder:2005hy,Chetyrkin:2005ia}, 
which is numerically negligible, 
so that $F_{2,h}^{ZMVFN}$ and $F_{2,h}^{asy}$ in Eq.~(\ref{eq:bmsn}) essentially cancel.
Therefore, at small $Q^2$ one obtains $F_{2,h}^{BMSN} \to F_{2,h}^{FFN}$.
On the other hand, at large scales $Q^2 \gg m_h^2$ the FFN term is canceled 
by $F_{2,h}^{asy}$ and one has in this limit, that $F_{2,h}^{BMSN} \to F_{2,h}^{ZMVFN}$. 
In summary, the BMSN prescription Eq.~(\ref{eq:bmsn}) provides a smooth 
transition between FFN scheme at small momentum transfer to the ZMVFN scheme at large 
scales, cf. Fig.~\ref{fig:bmsn}.
%
\begin{figure}[tbp]
\centering 
\includegraphics[width=.9\textwidth,height=0.8\textwidth]{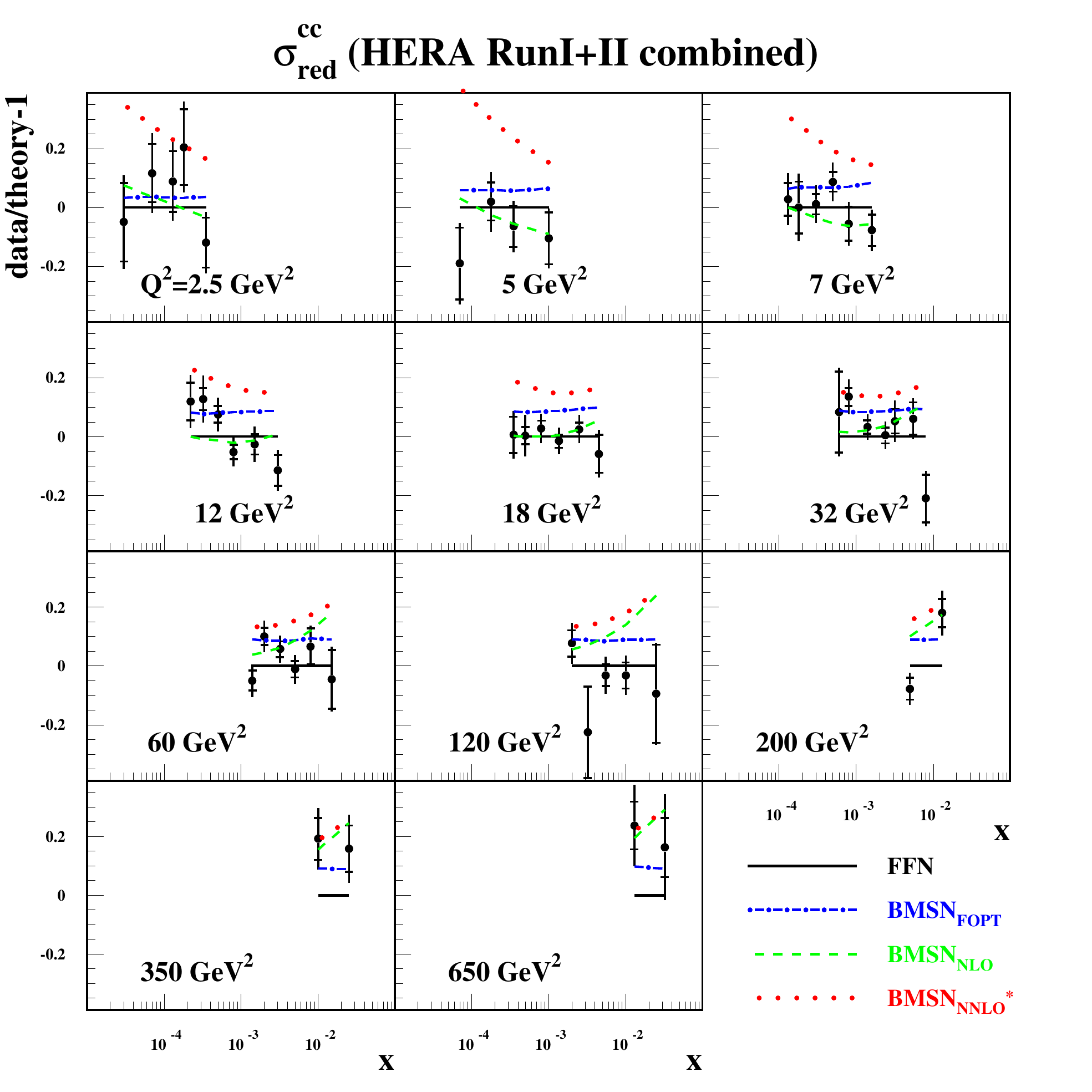}
\hfill
  \caption{\small 
    The pulls obtained for the combined HERA data 
    on DIS $c$-quark production~\cite{H1:2018flt} 
    in the FFN version of the present analysis (solid lines) versus $x$ in bins on $Q^2$.
    The predictions obtained using the BMSN prescriptions of the VFN scheme 
    with various versions of the heavy-quark PDFs with respect to 
    the FFN fit are displayed for comparison 
    (dotted-dashes: fixed order NLO; dashes: evolved from the NLO matching conditions 
    with the NLO splitting functions; 
    dots: the same for the NLO matching conditions combined with the NNLO splitting functions).  
    The PDFs obtained in the FFN fit are used throughout.
  }
  \label{fig:heracvfn}
\end{figure}

A version of the BMSN prescription based on the NLO evolution of the 
$(n_f+1)$-flavor PDFs also allows for a smooth matching with the FFN scheme 
at $Q^2 = m_h^2$, because the NLO-evolved PDFs do not have a discontinuity with respect to the FOPT ones
at the matching point, cf. Figs.~\ref{fig:pdfevol} and \ref{fig:f2c}. 
For the variant denoted NNLO$^{\ast}$ which uses NNLO-evolved PDFs the trend is different:
the slope of $F_{2,h}$ at $Q^2 = m_h^2$ predicted by the BMSN prescription is
substantially larger than the one obtained with the FFN scheme.
This is in line with the difference between the NNLO and FOPT PDFs. 
Obviously, this difference is not explained by the impact of the resummation of large logarithms, 
but rather by the mismatch in the perturbative order of the matching conditions 
and evolution kernels employed to obtain the NNLO heavy-quark PDFs. 
Therefore, the difference between the NLO and NNLO$^{\ast}$ variants 
of the VFN scheme should essentially quantify its uncertainty 
due to the missing NNLO corrections to the massive OMEs.
A choice of the matching scale $\mu_0=m_h$, i.e., at the mass of the heavy quark,  
is a matter of convention rather than appearing as a consequence of solid theoretical arguments.
Also note, that for DIS charm production, 
the matching scale $\mu_0$ cannot be significantly shifted to scales much lower than $m_c$, 
because in this case the matching would be performed at scales well below 1~GeV, 
where QCD perturbative expansions are not converging anymore.
When $\mu_0$ is shifted upwards, e.g., $\mu_0=2m_h$, the difference between
the NLO and NNLO$^{\ast}$ variants of the VFN scheme is becoming less significant.
This is particularly due to the fact, that then essential parts of the problematic small-$Q^2$ region  
are left for a theoretical description within the FFN scheme, cf. Fig.~\ref{fig:f2cth}.

The impact of scheme variations and the choice of the matching scale are 
qualitatively similar for the $c$- and $b$-quark production. 
Nonetheless, the effects are less pronounced for the $b$-quark case~\cite{Bertone:2017ehk},
mainly because of the smaller numerical value of strong coupling at the scale $m_b$. 
For this reason, and also due to more representative kinematics of data, all
our phenomenological comparisons are focused on the $c$-quark contribution.

\section{Benchmarking of the FFN and VFN schemes with the HERA data} 
\label{sec:pheno}

\begin{figure}
  \centering
  \includegraphics[width=0.455\textwidth,height=0.42\textwidth]{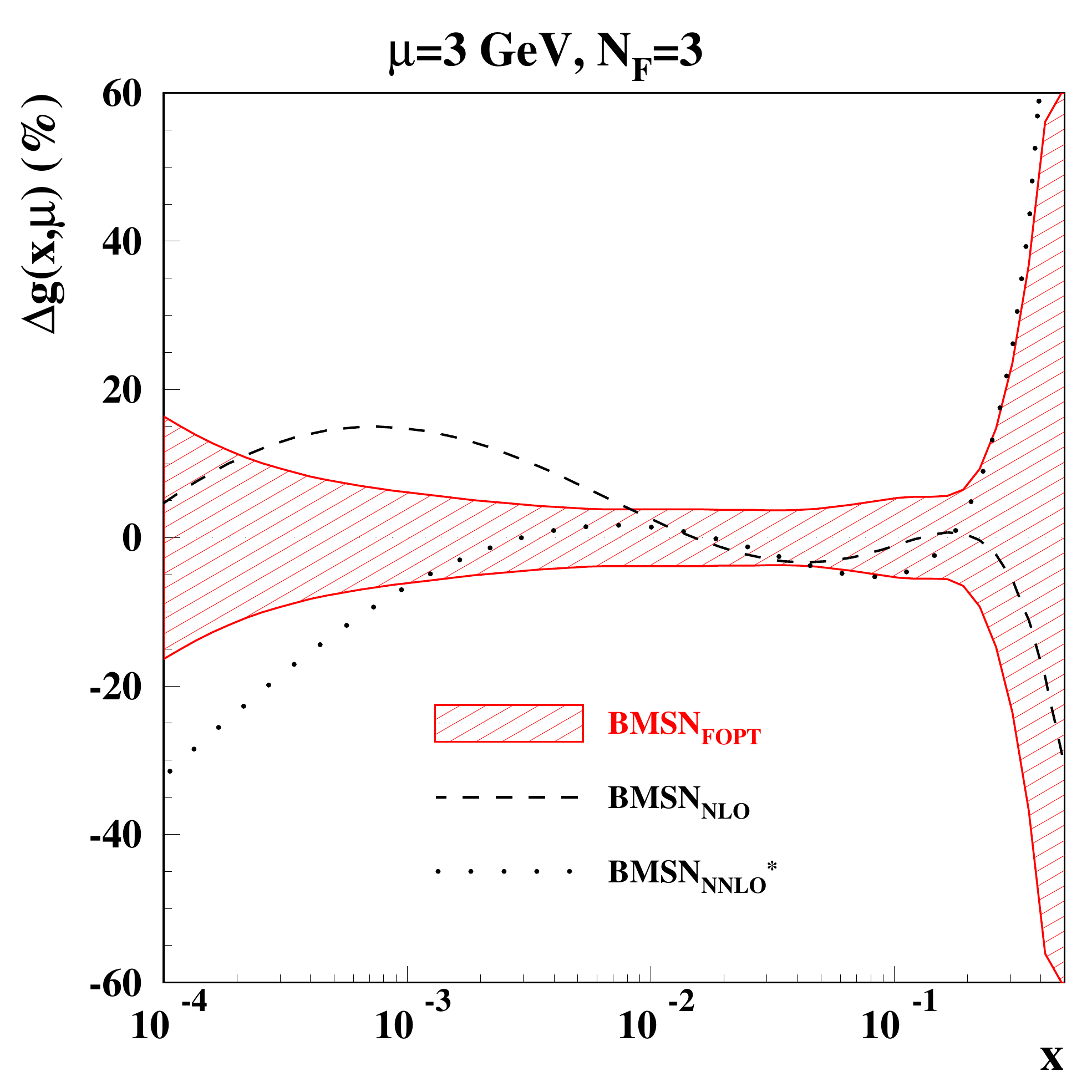}
  \includegraphics[width=0.455\textwidth,height=0.42\textwidth]{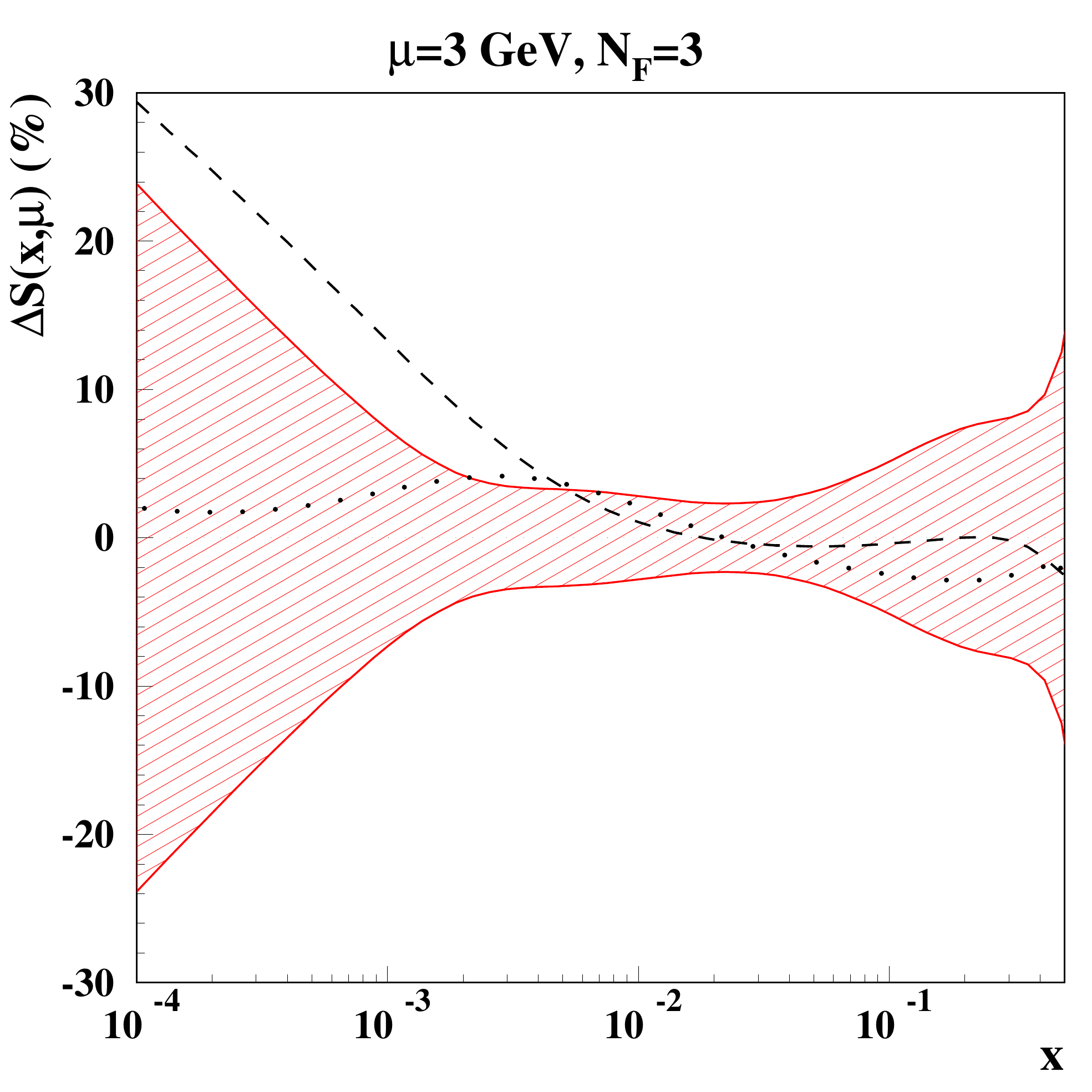}
  \caption{\small Left panel: 
The relative uncertainty in the 3-flavor gluon distribution $xg(x,\mu)$ 
at the factorization scale $\mu=3~{\rm GeV}$ versus $x$
obtained in the fit based on the BMSN VFN prescription with the 
FOPT heavy-flavor PDFs (hatched area) 
in comparison to the relative variation of its central value due to 
switching to the NLO- (dashes) and NNLO$^{\ast}$-evolved (dots)
PDFs. Right panel: the same for the total 
light-flavor sea quark distribution $xS(x,\mu)$. 
}
    \label{fig:pdfs}
\end{figure}
To study the phenomenological impact of the VFN scheme uncertainties 
we consider several variants of ABMP16 PDF fit~\cite{Alekhin:2017kpj}, 
which include the recent HERA data on heavy-flavor DIS production~\cite{H1:2018flt}.
Furthermore, the inclusive neutral-current DIS HERA data used in the 
ABMP16 fit are excluded in order to illuminate the impact of the scheme variation 
on the PDFs extracted from the fit. 
For the same reason we exclude the collider data on $W^\pm$- and $Z$-boson production, 
which provide an additional constraint at small-$x$ on the PDFs in the ABMP16 fit. 
However, in order to keep the different species of quark flavors disentangled, 
we add data on DIS off a deuteron target, analogous to an earlier study in Ref.~\cite{Alekhin:2017fpf}. 
For all variants we employ the NLO massive Wilson coefficients~\cite{Laenen:1992zk} 
and the pole-mass definition for the heavy-quark masses, so that a consistent 
comparison of the FFN scheme with the original formulation of the BMSN
prescription and its modifications is possible. For the same purpose we 
take the factorization scale $\mu_F={m_h}$ both for the FFN and VFN scheme.   
The values of $m_c^{pole}=1.4$~GeV and $m_b^{pole}=4.4$~GeV used in the present study are not perfectly 
consistent with the ones obtained in the ABMP16 fit with the \msbar
definition.
However, they are close to the values in the pole-mass scheme 
preferred by the HERA data~\cite{H1:2018flt}~\footnote{
Changing the heavy-quark mass renormalization scheme 
to the $\overline{\rm MS}$-scheme is straightforward, cf.~\cite{Alekhin:2010sv,Ablinger:2014nga}.}.
With these settings, the FFN scheme provides a good description of the $c$-quark production 
data, cf. Fig.~\ref{fig:heracvfn}. 
The agreement of the fit with the data is equally good, both at small and at large $Q^2$, 
underpinning the fact, that any additional large logarithms cannot improve 
the theoretical data treatment within the range of kinematics covered by the HERA data. 
This observation is indeed long known~\cite{Gluck:1993dpa}.

In order to check this aspect in greater detail we also compare 
predictions of various versions of the VFN scheme with the data. 
Let us consider the VFN predictions for the heavy-quark production cross sections 
which are computed by using the BMSN prescription of Eq.~(\ref{eq:bmsn}) for $F_2$, 
while still keeping the FFN scheme for $F_L$.
The justification of this approach derives from the small numerical contribution of $F_L$ 
as compared to $F_2$. 
In addition, the modeling of $F_L$ within the VFN framework is conceptually
problematic~\cite{Alekhin:2009ni}, because the effects of power 
corrections in $m_h^2/Q^2$ cannot be disregarded for this observable~\cite{Buza:1995ie}.
The PDFs used in this comparison are the ones obtained in
the FFN version of the fit. 
Therefore the obtained pulls display the impact of the scheme variation only. 

As expected, predictions of the VFN scheme based on the BMSN prescription and 
the FOPT heavy-flavor PDFs are close to the FFN ones. 
The same applies to the case of NLO-evolved PDFs, which are smoothly matched with the FFN ones
at small scales, cf. Fig.~\ref{fig:f2c}. 
In contrast, an excess with respect to the small-$Q^2$ data appears 
for the variant of the fit with the NNLO$^{\ast}$ PDFs employed. 
This excess is clearly related to the mismatch between the FFN scheme 
and this variant of the VFN one. 
At large $Q^2$ the impact of the resummation of large logarithms is marginal, 
in particular given the size of the data uncertainties. 
The latter is true also for the case of NLO-evolved PDFs.  

\begin{figure}
  \centering
  \includegraphics[width=0.455\textwidth,height=0.42\textwidth]{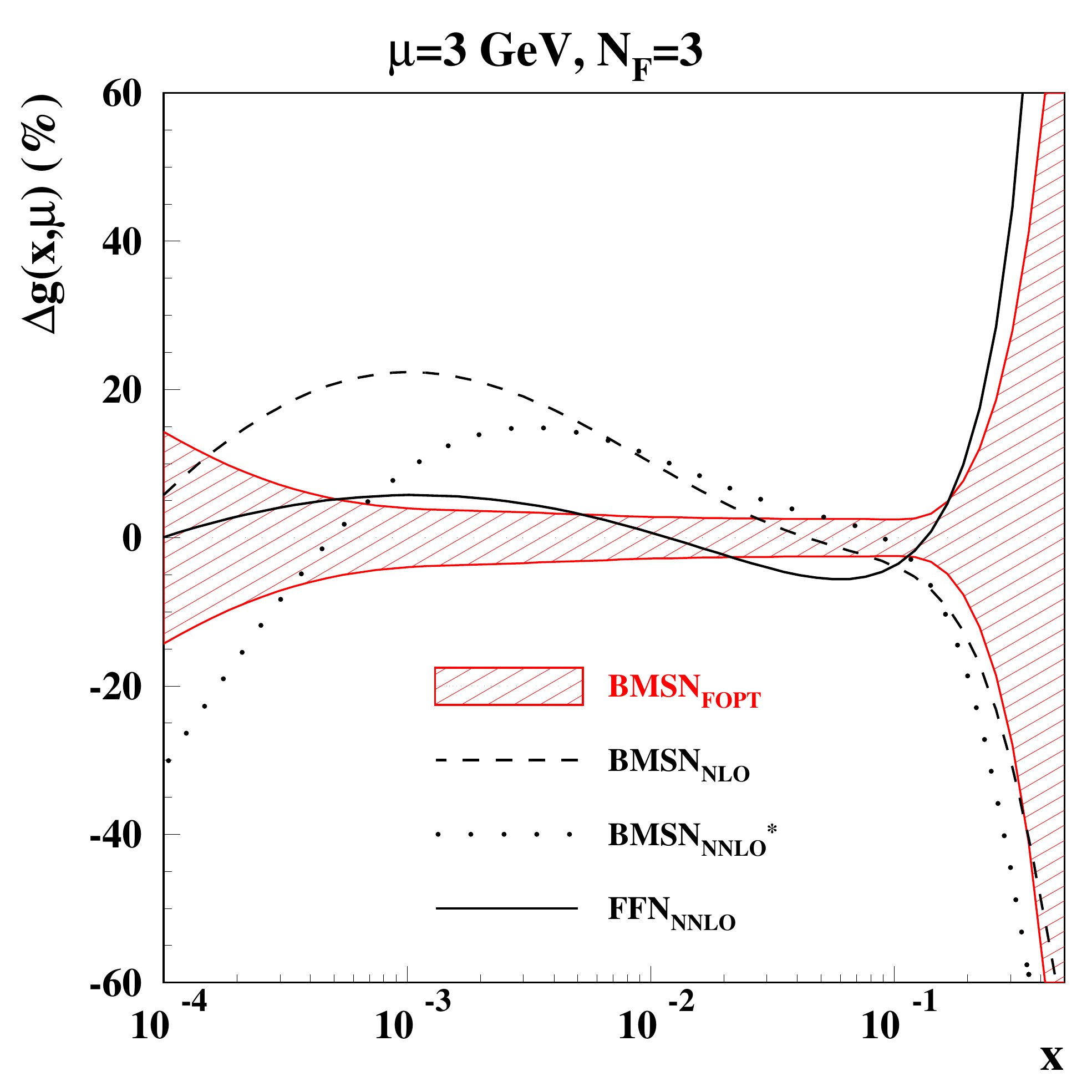}
  \includegraphics[width=0.455\textwidth,height=0.42\textwidth]{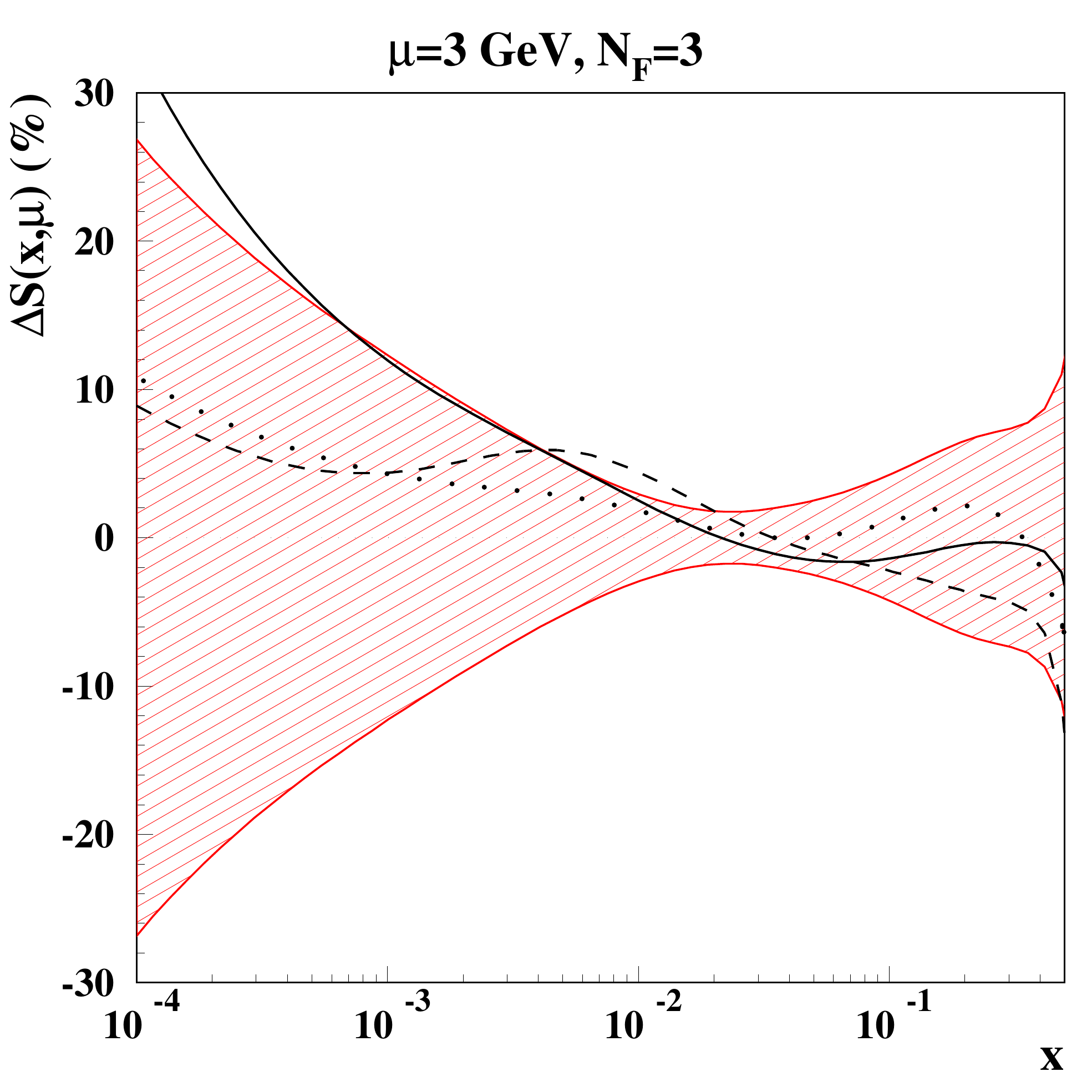}
  \caption{\small The same as in Fig.~\ref{fig:pdfs} for the variants of the fit 
    with the HERA inclusive DIS data appended. 
    Results of the NNLO FFN fit displayed for comparison (solid lines). 
  }
  \label{fig:pdfsin}
\end{figure}
\begin{figure}
  \centering
  \includegraphics[width=0.455\textwidth,height=0.42\textwidth]{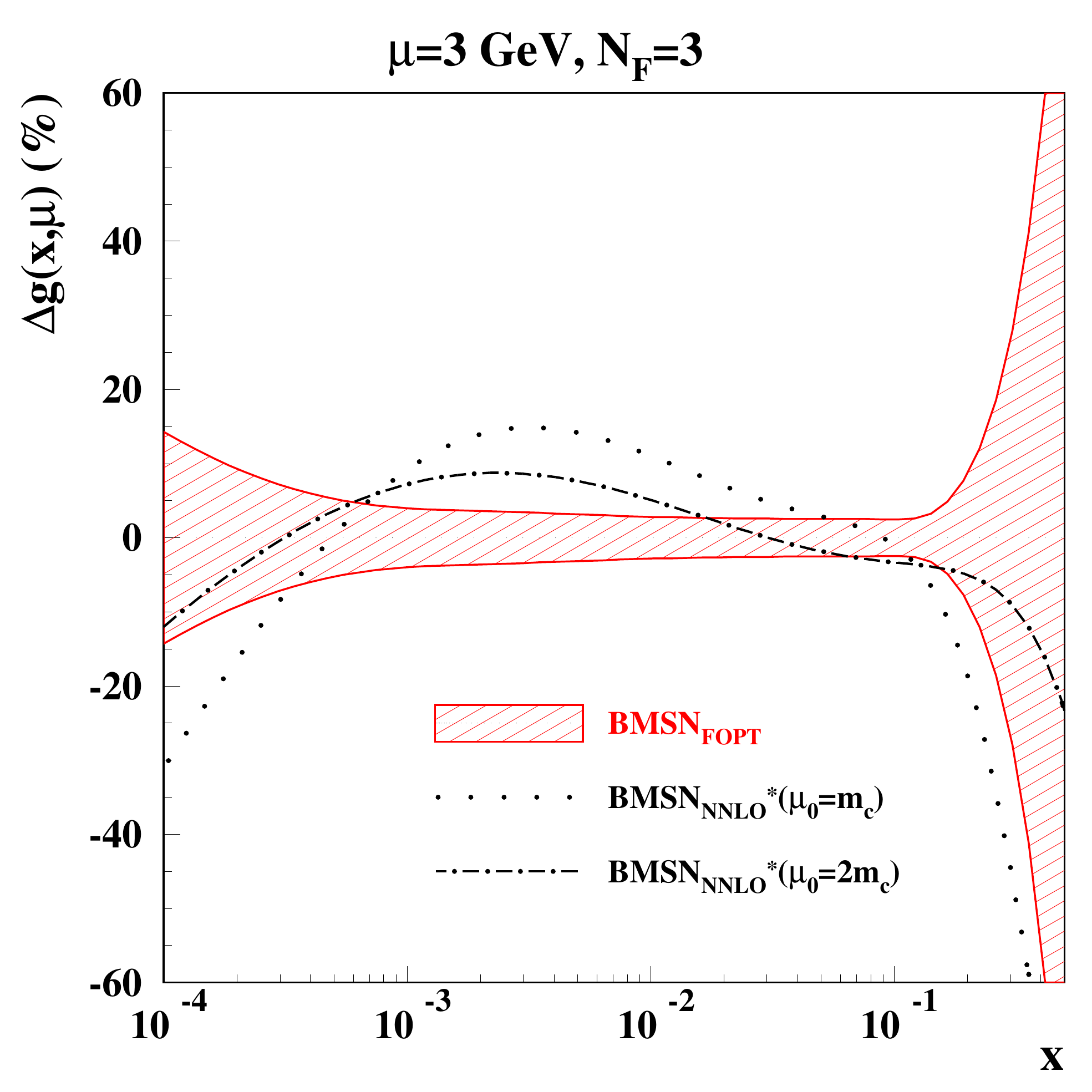}
  \includegraphics[width=0.455\textwidth,height=0.42\textwidth]{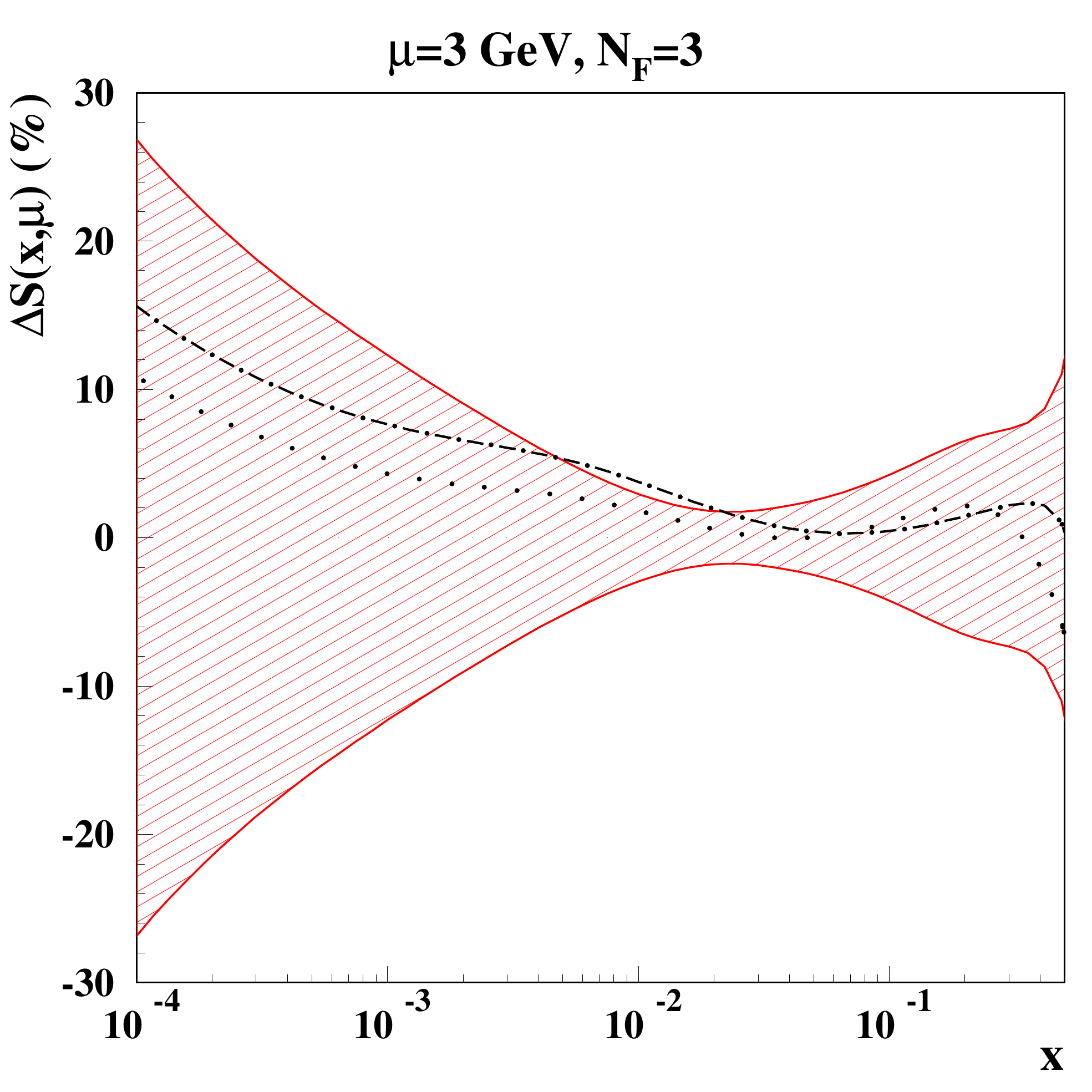}
  \caption{\small The same as in Fig.~\ref{fig:pdfsin} for a comparison  
    of two variants of the BMSN fit based on the NNLO$^{\ast}$ PDFs 
    with the matching point at $\mu_0=m_c$ (dots) and $\mu_0=2m_c$ (dashed dots).
  }
  \label{fig:pdfshi}
\end{figure}

The HERA data on $c$-quark production used in the present analysis are 
accurate enough to provide a sensible constraint on the small-$x$ gluon 
distribution. Moreover, the latter demonstrate 
sensitivity to the choice of the factorization scheme, cf. Fig.~\ref{fig:pdfs}.
The FFN scheme and the BMSN scheme with the FOPT and the NLO-evolved PDFs are in 
qualitative agreement, 
while a much lower small-$x$ gluon distribution is preferred
in the variant based on the NNLO$^{\ast}$ PDFs.
This is in line with the trends observed for the pull comparison, cf. Fig.~\ref{fig:heracvfn}. 

The difference between gluon and quark distributions obtained in the NLO- and NNLO$^{\ast}$-based fits 
is pronounced at small $x$ due to kinematic correlations with the small-$Q^2$ region, 
where the difference between these two approaches is localized, and reaches $\sim 30\%$ at $x=10^{-4}$.
The description of the small-$x$ inclusive DIS data is also sensitive to the 
scheme choice due to a substantial contribution of the heavy-quark production. 
In order to check this quantitatively, we consider variants of the fits with various VFN scheme prescriptions and the 
HERA inclusive data~\cite{Abramowicz:2015mha} added. 
In line with the recent update of the ABMP16 fit~\cite{Alekhin:2019ntu},
we impose strong cuts on the momentum transfer $Q^2 > 10~{\rm GeV}^2$ and 
on the hadronic mass  $W^2 > 12.5~{\rm GeV}^2$, 
which allow to avoid any impact of higher twist corrections, cf.~\cite{Alekhin:2012ig}.
The PDF uncertainties are improved due to the additional data included. 
However, the sensitivity of the resulting gluon distribution to the 
choice of heavy-quark PDF evolution still reaches $\sim30\%$ at $x=10^{-4}$, cf. Fig.~\ref{fig:pdfsin}. 
Such a spread induces quite essential uncertainties in the small-$x$ VFN predictions, 
in particular in the $c$- and $b$-quark input distributions 
for scattering processes at hadron colliders. 
\begin{figure}
  \centering
  \includegraphics[width=0.455\textwidth,height=0.42\textwidth]{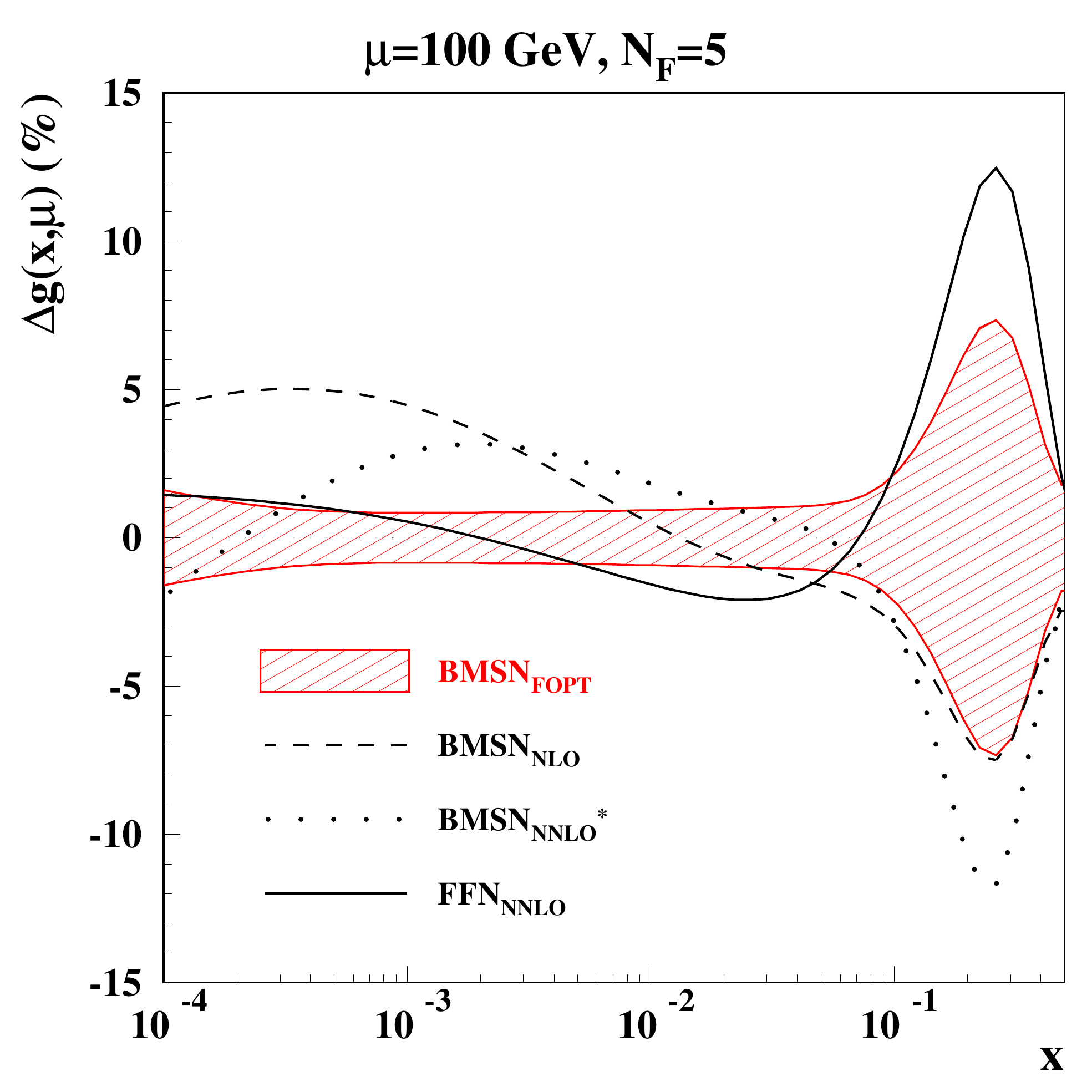}
  \includegraphics[width=0.455\textwidth,height=0.42\textwidth]{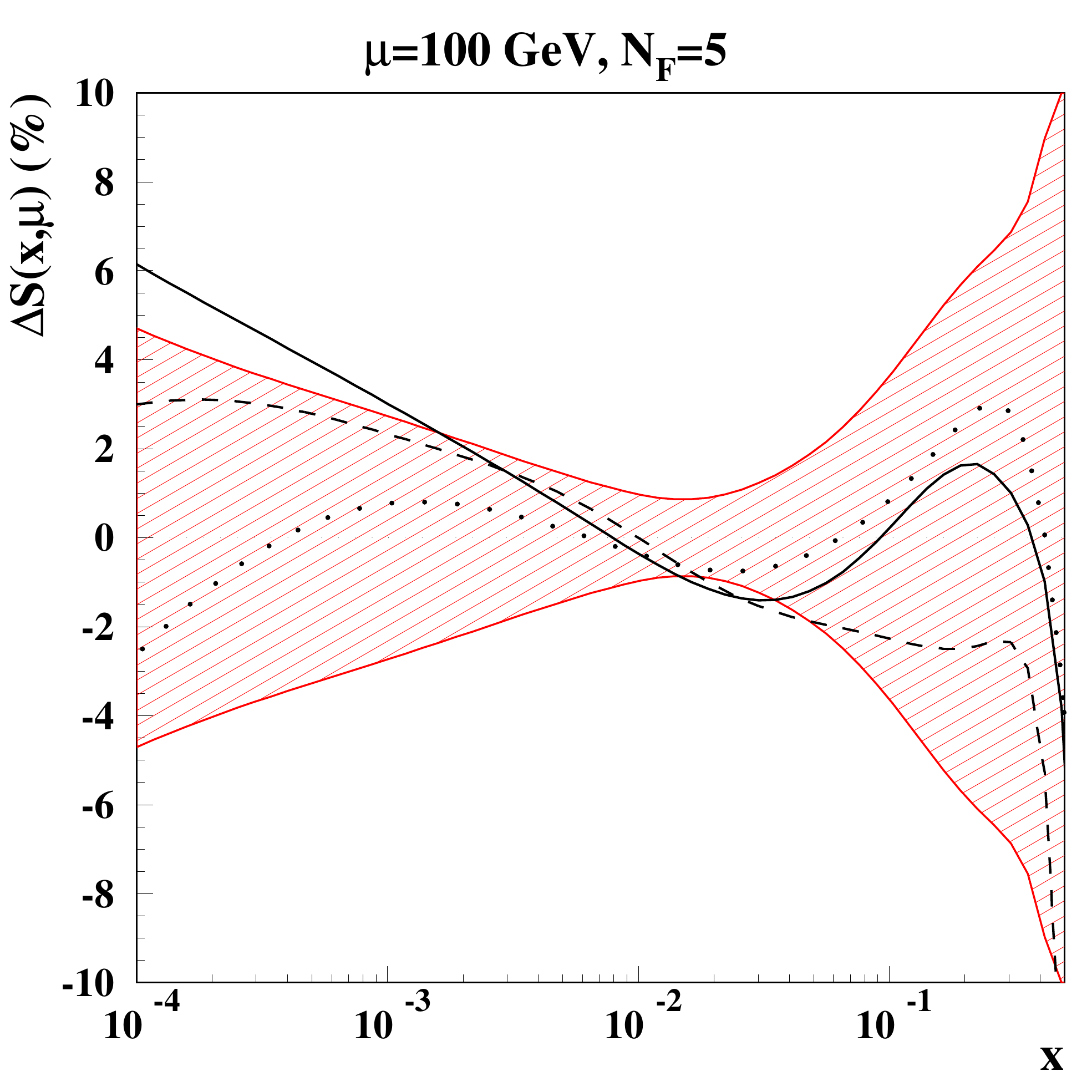}
  \caption{\small The same as in Fig.~\ref{fig:pdfsin} for the 5-flavor PDFs 
    at the factorization scale $\mu=100$~GeV. 
  }
  \label{fig:pdfhigh}
\end{figure}

The gluon distribution obtained using the BMSN  prescription 
with the NLO-evolved PDFs is increased with respect to 
the FOPT one at $x \sim 0.01$, which gives a hint on the impact of the
resummation of large logarithms at these kinematics. 
No further substantial change in the gluon distribution at $x \gtrsim 0.01$
is observed, when the NNLO corrections to the evolution are taken into account.
Therefore, one should expect a minor impact of 
the logarithmic terms at higher order (higher powers) 
on the description of the existing DIS data, although
the comparison is somewhat deteriorated by the uncertainty 
from the mismatch in perturbative orders in the NNLO$^{\ast}$ fits appearing at $x\lesssim 0.01$.
In this context it is also instructive to consider the results of the FFN 
fit performed with account of the NNLO corrections, which include 
the terms up to $O(\ln^2(\mu^2/m_h^2))$~\cite{Alekhin:2017kpj} and 
\msbar masses $m_c(m_c)=1.27~{\rm GeV}$, $m_b(m_b)=4.18~{\rm GeV}$~\cite{Tanabashi:2018oca}.
The gluon distribution obtained with these settings 
is similar to the VFN ones at $x\gtrsim 0.01$, 
located between the NLO and NNLO$^{\ast}$ fit results at $x\sim 10^{-4}$
and lower by $\sim 5\%$ than both of these variants at $x\sim 0.01$, 
where they agree with each other, cf. Fig.~\ref{fig:pdfsin}. 
This plot also yields an upper limit on the estimate of the impact of missing large logarithms in the NNLO FFN fit. 
On the other hand, a comparison with the NNLO VFN fit at small $x$ is inconclusive due to 
the large uncertainties in the VFN scheme appearing at these kinematics.
A more accurate estimate requires the NNLO VFN fit with a consistent boundary condition based on 
OMEs at NNLO accuracy~\cite{Bierenbaum:2009mv,Ablinger:2010ty,Kawamura:2012cr,Ablinger:2014lka,Ablinger:2014nga,Alekhin:2017kpj}.
Nonetheless, at the present level of data accuracy this upper limit is comparable with the experimental 
uncertainties in the gluon distribution obtained from the fit. 

Finally, considering a variation of the matching scale for the 4-flavor PDFs from $\mu_0=m_c$ to $\mu_0=2m_c$, 
leads to VFN heavy-flavor predictions being closer to the FFN ones, cf. Fig.~\ref{fig:f2cth}. 
The phenomenological effect of such a variation is more substantial at small $Q^2$ and $x$ due to
kinematic characteristics of the existing DIS experiments. 
Therefore, the corresponding change of the gluon distribution due to a matching point 
variation is significant mostly at $x \lesssim 10^{-3}$, cf. Fig.~\ref{fig:pdfshi}. 
It is comparable in size with the VFN scheme uncertainty related to the
boundary conditions for the evolution. 
However, strictly speaking, these two uncertainty sources should not be considered
independently since the impact of the matching scale variation also manifests itself 
through the scheme change. 

\section{Implications of VFN schemes for predictions at hadron colliders} 
\label{sec:colliders}

The contribution of heavy flavors to the hadro-production of massive states, 
like $W^\pm$-, $Z$- and  Higgs-bosons, $t$-quarks, etc., 
are commonly taken into account within the 4- or 5-flavor scheme.
This allows for great simplifications of the computations, 
since the VFN PDFs employed in this case contain resummation effects, 
which are generally rising with the factorization scale, cf.~Fig.~\ref{fig:pdfevol}. 
Therefore, the VFN scheme provides a relevant framework 
for the phenomenology of heavy particle hadro-production. 

The NNLO 4- and 5-flavor PDFs still suffer from the uncertainty due to the yet 
unknown exact NNLO corrections to the massive OMEs.
Moreover, for the NNLO PDFs derived from the VFN fit including the small-$x$ DIS data 
this uncertainty is enhanced, since the part of those DIS data, 
which provides an essential constraint on the PDFs, also populates the matching region. 
The observed spread in the 5-flavor gluon distributions, which are obtained
from the VFN fits with varying treatments of the matching ambiguity 
is somewhat reduced with increasing scales due to the general properties of the QCD evolution. 
However, it is still comparable to experimental uncertainties at $x\sim 0.01$ and 
substantially larger at $x\sim 10^{-4}$, cf.~Fig.~\ref{fig:pdfhigh}.

Altogether, this implies an uncertainty in predictions of the production rates of 
the Higgs-boson and $t$-quark pairs at the Large Hadron Collider (LHC) within a margin of few percent and 
somewhat larger at the higher collision energies discussed for future hadron colliders. 
Note that in the ABMP16 fit~\cite{Alekhin:2017kpj}, which is based on a
combination of both, DIS and hadron collider data, 
the FFN and the 5-flavor VFN schemes are used 
for the theoretical description of these samples, respectively.
This allows to keep the advantages of the VFN scheme at large scales, 
while avoiding its problems concerning the DIS data. 
Nevertheless, the NNLO massive OMEs~\cite{Bierenbaum:2009mv,Ablinger:2010ty,Kawamura:2012cr,Ablinger:2014lka,Ablinger:2014nga,Alekhin:2017kpj} 
are still necessary 
to generate NNLO PDFs free from the matching ambiguity.

\begin{figure}
  \centering
  \includegraphics[width=0.9\textwidth,height=0.42\textwidth]{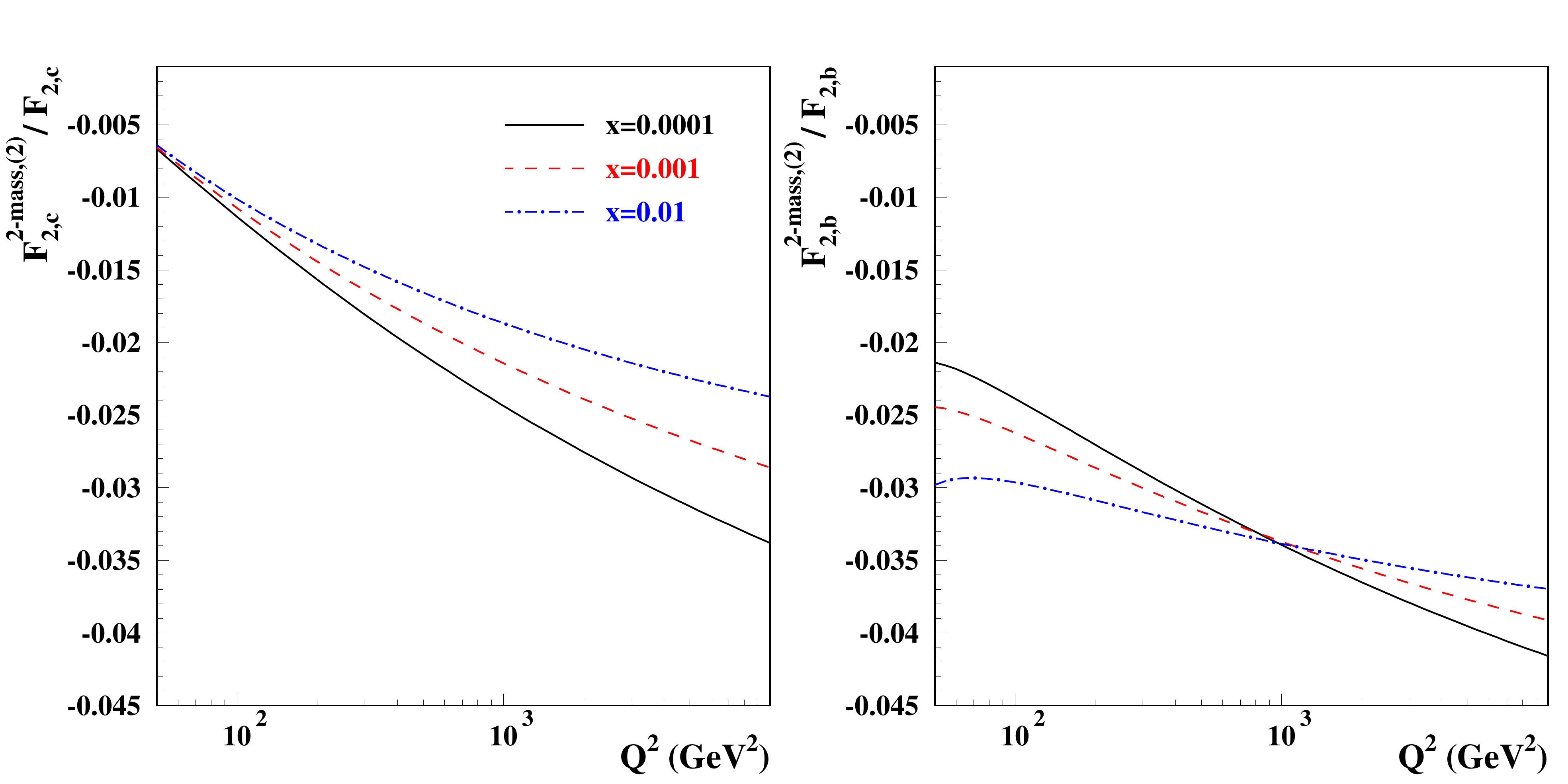}
  \caption{\small 
The ratio of two-mass contribution Eq.~(\ref{eq:tm}) to the DIS
    structure function $F_{2,c}$ (left panel) and $F_{2,b}$ (right panel)
computed in the VFN scheme using the PDFs from the NNLO$^{\ast}$ variant of the VFN fit
versus momentum transfer $Q^2$ and at various values of Bjorken $x$ (solid line: 
$x$=0.0001, dashes: $x$=0.001, dotted-dashes: $x$=0.01). 
  }
  \label{fig:tm}
\end{figure}
%
In closing the studies of VFN schemes we wish to address a conceptual problem
of the 5-flavor scheme definition due to the fact, 
that the $b$-quark mass $m_b$ is not too much larger than $m_c$.
This relates to the inherent limitations of the VFN schemes due to the successive decoupling 
of one heavy quark at a given time. 
As discussed above, starting from the two-loop order, the DIS structure functions also receive contributions 
which contain two different massive quarks~\cite{Blumlein:2018jfm}. 
At two loops, they are given by one-particle reducible Feynman diagrams,
while one-particle irreducible graphs appear at the three-loop order for the first
time, cf.~\cite{Ablinger:2017err,Ablinger:2017xml,Ablinger:2018brx}.

Here we will consider the two-loop effects, which arise from virtual corrections 
with both, charm and bottom quarks.
Thus, no production threshold is involved. 
For the structure function $F_2$ one obtains 
\begin{eqnarray}
\label{eq:tm}
F_{2,h}^{{\rm 2-mass},(2)}(x,Q^2) \,=\, - e_h^2\
a_s^2(Q^2)\, \frac{16}{3} T_F^2\, x\, \ln\left(\frac{Q^2}{m_c^2}\right) \ln\left(\frac{Q^2}{m_b^2}\right)
\int\limits_x^1 \frac{dz}{z} \left(z^2 + (1-z)^2\right)\, g\left(\frac{x}{z},Q^2\right)\, ,
\end{eqnarray}
which is to be added to Eqs.~(\ref{eq:zmvfn}) or (\ref{eq:ffn}), 
with $e_h$ denoting the fractional heavy-quark charge and using $T_F =1/2$.
The effect of the 2-mass contributions rises at small $x$ and large $Q^2$, being more pronounced for the case 
of 
$b$-quark production, cf. Fig.~\ref{fig:tm}. For the kinematics of the proposed lepton-proton LHeC collider it 
reaches 
up to $\sim 3\%$, which has impact on the phenomenology of heavy-quark production.

As demonstrated in Ref.~\cite{Blumlein:2018jfm}, 
the two-mass diagrams at the two-loop order have the largest effects 
for the $b$- and $c$-quark distribution at large $Q^2$.
The respective PDFs can be obtained by adding the two-mass contributions to the OMEs in Eq.~(\ref{eq:VFNS-hq}).
Comparing the heavy-quark PDFs with and without the two-mass effects included,
one finds that the relative size of the effect is negative: 
$b$-quark distributions with the two-mass contributions included are decreased
by -2\% to -6\% in the range for $Q^2$ from 30 to $10000~\GeV^2$ at small $x$, $x=10^{-4}$;
$c$-quark distribution the relative variations are smaller, amounting to -1\% to -4\% 
for $Q^2 = 100~\GeV^2$ to $10000~\GeV^2$ and $x=10^{-4}$.
In precision fits these two-mass effects have consequences for all PDFs and 
require the use of a different VFN scheme 
compared to those with the decoupling of a single heavy quark at the time, cf.~\cite{Blumlein:2018jfm}. 
At this point, however, we leave detailed studies of VFN schemes with two
massive quarks, i.e., the simultaneous transition $f_i(n_f) \to f_i(n_f+2)$
for PDFs for future studies.

\section{Conclusions}
\label{sec:concl}

The precise description of the parton content in the proton across a large
range of scales is a an important ingredient in precision phenomenology.
The treatment of heavy quarks with a mass $m_h$ requires adapting the 
number of light flavors in QCD to the kinematics under consideration,  
set by the factorization scale $\mu$, which is typically
associated with the hard scale of the scattering process.
Within the ABMP16 global PDF fit, the FFN scheme with $n_f=3$ light flavors 
provides a good description of the existing world DIS data, while
the LHC processes are typically described with $n_f=5$ massless flavors by 
implementing decoupling of heavy quarks and a
transition from 3- to 4- or 5-flavor PDFs, 
including the possibility for the resummation of large logarithms in $Q^2/m_h^2$.

To check the effects of such a resummation on the analysis of existing DIS data 
we have studied the $c$-quark PDF, constructed with the help of massive OMEs in QCD, 
and we have quantified differences between the use of perturbation theory at fixed
order and subsequent evolution.
We have found that the impact of the PDF evolution as used in the BMSN prescription of VFN scheme 
is sizable and rather $x$-dependent than $Q^2$-dependent, showing little impact 
on the large-log resummation on the heavy-quark production at realistic kinematics. 
Moreover, these differences must be considered an inherent theoretical uncertainty of VFN schemes since
using NLO or NNLO accuracy for the evolution leads to significantly different results due  to 
mismatch in the orders of perturbation theory between the heavy-quark OMEs  
and the accuracy of the evolution equations.
Likewise, and related, the choice of the matching point position employed 
in the VFN schemes has the impact on heavy-quark PDFs and therefore brings additional uncertainty.

With the help of variants of the ABMP16 PDF fit, we have confronted 
the FFN scheme and different realizations of VFN schemes (FOPT, evolved at NLO, evolved at
NNLO) in the BMSN approach with the combined HERA data and DIS $c$-quark production.
The FFN scheme delivers a very good description of those data and we have 
found little need for the additional resummation of large logarithms in the kinematic range covered by HERA.
From the fit variants, we have also determined the gluon and the total
light-flavor sea quark distributions, illustrating again the sizable numerical differences, 
obtained by adopting the respective VFN scheme variants.
Depending on the value of $x$, the observed differences for the gluon PDF 
are well outside the experimental uncertainties at low factorization scales 
and persist as well as at high scales of ${\cal O}(\text{100})~\GeV$.
The VFN scheme choices are, therefore, highly relevant for LHC phenomenology and affect the
predictions for the hadro-production of massive particles within a margin of few percent.

In summary, despite being applicable in a limited kinematic range, 
the FFN scheme works very well for the modern PDF fits 
and contains much smaller theoretical uncertainty than the VFN schemes currently available. 
As an avenue of future development, 
the latter will benefit from improving the perturbative accuracy of the 
massive OMEs used, including their NNLO corrections, which are known exactly 
or to a good approximation. 
Other features of VFN schemes to be improved concern the simultaneous 
decoupling of bottom and charm quarks, which is advisable 
due to the close proximity of the mass scales $m_b$ and $m_c$. 
We leave these issues for future studies.

\acknowledgments
\noindent
This work has been supported in part by  by Bundesministerium f\"ur Bildung und Forschung (contract 05H18GUCC1)
and by the EU ITN network SAGEX agreement No. 764850 (Marie Sk\l{}odowska-Curie).

\appendix
\renewcommand{\theequation}{\ref{sec:appA}.\arabic{equation}}
\setcounter{equation}{0}
\section{Heavy-quark scheme implementations}
\label{sec:appA}

We briefly summarize the technical details of the various implementations 
of heavy-quark schemes in PDF fits of 
ABMP16~\cite{Alekhin:2017kpj}, CT18~\cite{Hou:2019efy},
MMHT14~\cite{Harland-Lang:2014zoa} and NNPDF3.1~\cite{Ball:2017nwa}.

\subsection*{ABM} 
\noindent

The ABMP16 PDF fit~\cite{Alekhin:2017kpj} is based on the FFN scheme in a part concerning 
heavy-flavor DIS production. Nonetheless, for the collider data on $t$-quark, 
$W$- and $Z$-boson production, where the VFN scheme is more relevant, the 5-flavor PDFs are constructed 
from the 3-flavor ones, see Eqs.~(\ref{eq:VFNS-hq})--(\ref{eq:VFNS-g}),
using currently available information on the heavy-quark OMEs and employing NNLO evolution 
for the matched PDFs. 
All relevant formulae are implemented in the code {\tt OPENQCDRAD} (version 2.1), 
which is publicly available~\cite{openqcdrad}.

\subsection*{CT}
CT18~\cite{Hou:2019efy} uses the ACOT 
VFN scheme~\cite{Aivazis:1993pi,Kramer:2000hn,Tung:2001mv}, 
specifically an NNLO realization~\cite{Guzzi:2011ew} of the so-called S-ACOT-$\chi$ variant. 
The S-ACOT-$\chi$ VFN scheme features a slow rescaling of the
parton momentum fractions $z$ 
in the argument of the respective massless Wilson coefficient functions in $F_{2,h}^{ZMVFN}$ in Eq.~(\ref{eq:zmvfn})
by replacing $z \to \chi = z \left( 1 + \frac{4m_h^2}{Q^2} \right)$, 
and restricting the integration range of $z$ in the convolutions 
to $x \left( 1 + \frac{4m_h^2}{Q^2} \right) \leq z \leq 1$ with the Bjorken variable $x$.
The slow rescaling is motivated by its properties to model energy conservation
in the DIS production of heavy final states.
Ref.~\cite{Guzzi:2011ew} also explores a wider family of rescaling choices, 
which interpolate smoothly between $z$ and $\chi$.

\subsection*{MSTW}
MMHT14~\cite{Harland-Lang:2014zoa} uses the RT VFN scheme~\cite{Thorne:1997ga}, 
specifically the TR' prescription from Ref.~\cite{Thorne:2006qt} for PDF fits at NNLO.
The RT scheme requires as a constraint the continuity of physical observables in the threshold region, i.e., 
for the expression for $F_{2,h}^{FFN}$ in Eq.~(\ref{eq:ffn}) below and 
$F_{2,h}^{ZMVFN}$ in Eqs.~(\ref{eq:zmvfn}) above threshold.
To that end, the derivative of the structure function, $dF_2/d\ln Q^2$ is supposed to be continuous at the
matching point $Q^2 = m_h^2$ in the gluon sector.
To achieve this modeling constraint, a $Q^2$-independent term is added above the matching
point to the expression for $F_{2,h}^{ZMVFN}$ to maintain continuity of the structure function.
The TR' prescription specifies this procedure up to NNLO~\cite{Thorne:2006qt}.

\subsection*{NNPDF}
NNPDF3.1~\cite{Ball:2017nwa} uses the FONLL VFN scheme~\cite{Forte:2010ta}, 
which has been devised to combine the heavy-quark DIS structure functions
and the ZMVFN expressions in analogy to Eq.~(\ref{eq:bmsn}).
FONLL suppresses the difference of $F_{2,h}^{ZMVFN}$ in Eq.~(\ref{eq:zmvfn}) and
the necessary subtraction term, i.e., the expression analogous to $F_{2,h}^{asy}$ in Eq.~(\ref{eq:asymp}),
which is needed to avoid double counting, 
with a kinematical damping factor $\left( 1 - \frac{Q^2}{m^2} \right)^2$.
In this manner, it is guaranteed, that only $F_{2,h}^{FFN}$ of Eq.~(\ref{eq:ffn})
remains for virtualities $Q^2 \simeq m_h^2$ near threshold. 
The variant FONLL-C is used to determine the PDFs at NNLO~\cite{Forte:2010ta}.


\end{document}